\documentclass[%
 reprint,
 amsmath,amssymb,
 aps,
floatfix,longbibliography
]{revtex4-2}

\usepackage{graphicx}
\usepackage{dcolumn}
\usepackage{bm}
\renewcommand{\u}{\mathbf{u}}
\newcommand{\De}{\textrm{De}}
\newcommand{\Wi}{\textrm{Wi}}
\newcommand{\sig}{\boldsymbol{\sigma}}
\newcommand{\tr}{\textrm{tr}}
\newcommand{\tauB}{\boldsymbol{\tau}}

\begin{document}

\title{Undulatory swimming in viscoelastic fluids under confinement}
\author{David A. Gagnon}
\affiliation{Department of Physics,
Georgetown University, Washington, DC 20057}
\affiliation{Institute for Soft Matter Synthesis and Metrology,
Georgetown University, Washington, DC 20057}

\author{Becca Thomases}
\affiliation{Department of Mathematical Sciences,
Smith College, Northampton, MA 01063}

\author{Robert D. Guy}
\affiliation{Department of Mathematics,
University of California Davis, Davis, CA 95616}

\author{Paulo E. Arratia}
\email{parratia@seas.upenn.edu}
\affiliation{Department of Mechanical Engineering and Applied Mechanics,
University of Pennsylvania, Philadelphia, PA 19104}

\date{\today}

\begin{abstract}

Low Reynolds number swimmers frequently move near boundaries, such as spirochetes moving through porous tissues and sperm navigating the reproductive tract. Furthermore, these microorganisms must often navigate non-Newtonian fluids such as mucus, which are typically shear-thinning and viscoelastic. Here, we experimentally investigate such a system using the model biological organism \textit{C. elegans} swimming through microfluidic channels containing viscous Newtonian fluids and viscoelastic fluids. Swimmer kinematics and resulting flow fields are measured as a function of channel width and therefore the strength of confinement. Results show that, for viscoelastic fluids, weak or moderate confinement can lead to enhancement in propulsion speed but for strong confinement this enhancement is lost and the swimming speed is slower than for an unconfined nematode. 
 We use theory developed for bending elastic filaments in viscoelastic fluids to show that while (weak) confinement leads to increases in swimming speed there is a, $\De-$ dependent, $\Wi$ (Weissenberg number) number transition from a linear stress response regime to a nonlinear (or exponential) stress response regime. The experimentally obtained velocity fields are used to calculate a Weissenberg number to show that the decrease in swimming speed with confinement is likely related to growth in elastic stresses around the swimmer. 

\end{abstract}

\maketitle

\section{Introduction}

Propulsion mechanisms at small length scales are often governed by surface forces and characterized by low-Reynolds-number hydrodynamics \cite{Lauga2009, Spagnolie2015}; in this regime, viscous linear forces dominate inertial ones \cite{Lauga2009}. In the absence of inertia, a microorganism, such as \textit{Escherichia coli} and \textit{Caenorhabditis elegans}, swims by deforming its body and therefore the solid-liquid interface at its surface~\cite{Elfring2015, Sznitman2010PoF}. The mechanics of swimming at low Re have been well-studied for more than 70 years, particularly in Newtonian fluids\cite{Taylor1951, Lighthill1976, Korta2007, Lauga2009, Guasto2010, Padmanabhan2012, Bilbao2013}. Many microorganisms, however, swim or move in fluids that contain polymers, particles, and/or large proteins~\cite{Spagnolie_ARFM_2023,Arratia_PRF2022_Complex,Spagnolie2015,Fauci2006,Lauga2007,Patteson2016,Shen2011}. These complex fluids typically display non-Newtonian rheology such as shear-rate dependent viscosity and viscoelasticity~\cite{Larson1999}. Recent studies have explored the effects of fluid elasticity and local fluid structure on propulsion speed and kinematics in idealized models and living organisms~\cite{Lauga2007, Fu2009, Leshansky2009, Fu2010, Teran2010, Juarez2010, Shen2011, Liu2011, Harman2012, Gagnon2013,Gagnon2014FIP, Patteson2015, Qin2015,martinez_pnas_2014,Thomases2014,Thomases2019, Arratia_PRF2022_Complex, Spagnolie_ARFM_2023}. It has become increasingly clear that fluid elastic stresses can significantly modify the swimming speed and kinematics of microorganisms. Whether swimming speed is increased or decreased, however, is highly dependent on the swimers' gait and biomechanical properties, and their coupling with the material properties of the fluid~\cite{Thomases2014,Riley_2014, Sznitman2010BJ}. 

\begin{figure}[!b]
\centerline{\includegraphics[width=.5\textwidth]{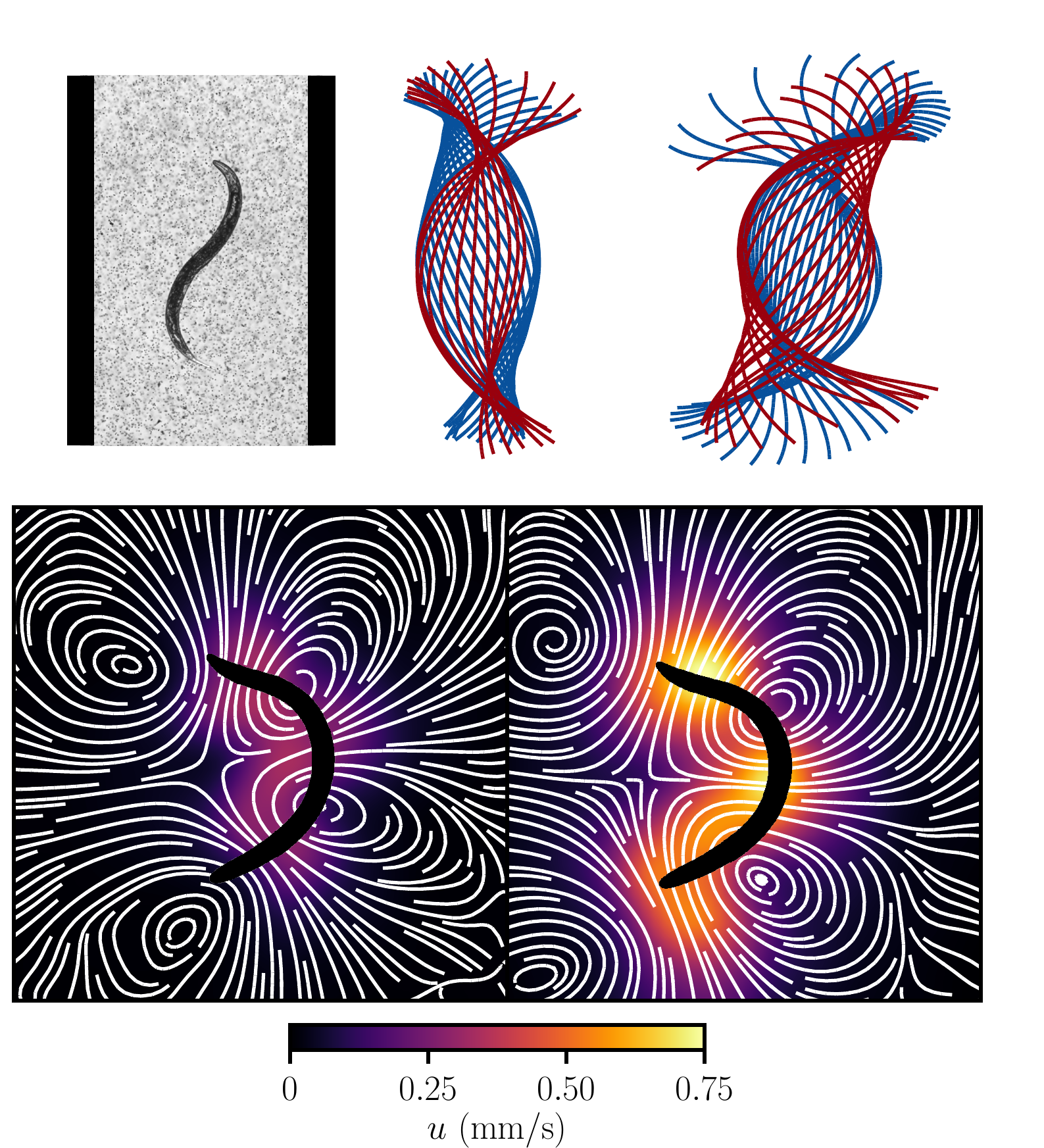}}
\caption{(Color available online) \textit{(a)} Schematic for \textit{C. elegans} swimming through a narrow channel of width $w$, producing transverse confinement. The fluid medium contains with 3.1~$\mu$m tracer particles. \textit{(b)} Nematode body shapes during one beating cycle in a viscosified Newtonian fluid under confinement. \textit{(c)} Nematode body shapes during one beating cycle in a viscoelastic fluid under confinement. \textit{(d)} Snapshot of the streamlines around an unconfined \textit{C. elegans} in a viscoelastic fluid; color represents fluid speed \textit{(e)} Streamlines and fluid speed around a confined \textit{C. elegans} in a viscoelastic fluid. Note that \textit{(d)} and \textit{(e)} are at $De \approx 1.2$.}
\label{Setup}
\end{figure}

In addition, many relevant biological processes rely on boundary interactions (walls, interfaces, etc) to function properly, which in turn can affect the swimmers' motility behavior \cite{Stone_2006, Wilding2021_PRL, Tokarova2021_PNAS}. Examples include sperm navigating confined fluid-solid interfaces \cite{Colburn1986,tung_PNAS_2015,Tung_PRL2015,suarez_sperm_2016} and spirochetes penetrating porous tissues~\cite{Harman2012}. A considerable amount of work has suggested surfaces significantly modify the behavior of swimmers. For example, early analytical studies~\cite{Taylor1951, Katz1974} and a more recent numerical simulations~\cite{Munch2016} using an infinite waving sheet in the presence of solid boundaries have shown that the sheet should swim faster under confinement in a Newtonian fluid. Similar results were found with a waving cylinder moving inside a capillary~\cite{Taylor1952, Felderhof2010}. Several studies have shown that hydrodynamic wall interactions can lead to a modification of drag forces~\cite{Evans2010}, swimmer aggregation in Newtonian fluids~\cite{Li2014}, and modified aggregation in complex fluids~\cite{Yazdi2014,Yazdi2015}. Additionally, work with \textit{C. elegans} has shown that the presence of solid boundaries can lead to a modulated swimming gait including a decrease in beating amplitude in Newtonian fluids~\cite{Schulman2014}. 

Despite the prevalence in nature and human physiology \cite{bansil_ARPhys,gaffney11}, the effects of boundaries on the swimming behavior of microorganisms in viscoelastic fluids has received considerably less attention. Using the squirmer model, simulations with the Giesekus fluid model showed that elastic stresses can trap "pusher" swimmers near walls \cite{Li2014,Ardekani_RheoActa_2014}. Results with Oldroyd-b model, a purely elastic model, seem to suggest that both "puller" and "pusher" swimmers are attracted to solid walls in viscoelastic fluids \cite{Yazdi2014}. Later, Elfring and Lauga \cite{Elfring_book2015} showed that, for a infinite (Taylor) waving sheet in a viscoelastic fluid near a wall, the small-amplitude sheet (or swimmer) speed decreases (relative to Newtonian fluids) due to the combined effects of viscoelasticity and confinement. Such a formulation is valid for prescribed kinematics (i.e., fixed frequency and amplitude) and builds on previous results of the waving sheet in viscoelastic media \cite{Lauga2007} to incorporate boundary effects.  Ives and Morozov \cite{Morozov_PoF2017} extended the analysis to finite amplitude swimmers using numerical simulations and found, surprisingly, that the waving sheet swimming speed can be higher than the corresponding Newtonian case; i.e., elastic stresses may enhance the speed of an undulatory swimmer near a wall. These results, however, have yet to be tested in experiments. 

Here, we experimentally investigate the combined effects of fluid elasticity and boundary confinement on the motility behavior of an undulatory swimmer. Experiments are performed using a model biological organism, the nematode \textit{C. elegans}. The nematode is observed swimming through narrow channels in polymeric fluids with varying levels of (fluid) elasticity and of (wall) confinement. We find that \textit{C. elegans} kinematics (e.g., swimming amplitude) are affected by its proximity to the solid boundaries. Our results show that the nematode's swimming speed decreases with increasing Deborah number ($De$) as the channel size becomes smaller than the wavelength of the nematode's sinusoidal beating.  However, the nematode's swimming speed is non-monotonic as a function of confinement or channel width at a particular $De$. We use the experimentally obtained velocity fields to calculate the flow Weissenberg number to show that the decrease in swimming speed with confinement occurs due to growth in elastic stresses around the swimmer.

\section{Methods}

\subsection{\textit{C. elegans}: a model organism for swimming studies}
\textit{C. elegans} are roundworms commonly used in studies of sleep, disease, and aging among many others. They are 1 mm in length and 80~$\mu$m in diameter, and their predominately planar sinusoidal swimming gait makes them an ideal model organism for studies of locomotion at low $Re$. We perform experiments in thin, fluid-filled acrylic channels 30 mm long and 3.125 mm deep and covered by a thin glass microscope cover slip (Fig.~\ref{Setup}\textit{(a)}). The width of the channels ranges from 0.8 mm to 2.5 mm. The smallest channel width is set by the maximum amplitude of the nematode's swimming gait, which is approximately 0.25 mm such that there is no direct contact with the wall. Additionally, we perform experiments in which nematodes are laterally unconfined using circular chambers of 2 cm diameter and 1 mm depth. 

Nematode's swimming motion is imaged via standard bright-field microscopy using an Infinity K2/SC microscope with an in-system amplifier, a CF-3 objective, and an IO Industries Flare M180 camera at 150 frames per second. The depth of focus of the objective is approximately 20~$\mu$m and the focal plane is set on the longitudinal axis of the nematode body. The nematode beats primarily in the observation plane; the out-of-plane beating amplitude of \textit{C. elegans} is less than 6\% of the amplitude of its in-plane motion~\cite{Sznitman2010PoF}. All data presented here pertain to nematodes swimming at the center of the fluidic chamber and out-of-plane recordings are discarded to avoid out-of-plane nematode-wall interactions. Ideal recordings are of the nematode swimming parallel to the channel walls and in the geometric center of the channel. Consequently, recordings where a nematode directly interacts with the wall or where a nematode swims at an angle with the wall greater than 15 degrees are also discarded. On average, 13 individuals are recorded for each combination of Newtonian or viscoelastic fluid and channel width.

We compute nematodes' swimming kinematics from videos using in-house software~\cite{RSznitman2010}. The software extracts the nematode's centroid position and body shape-line, and computes quantities such as swimming speed $U$, frequency $f$, wavelength $\Lambda$, and amplitude $A$.

\subsection{Fluid Properties}

\begin{figure}[t]
\centerline{\includegraphics[width=.5\textwidth]{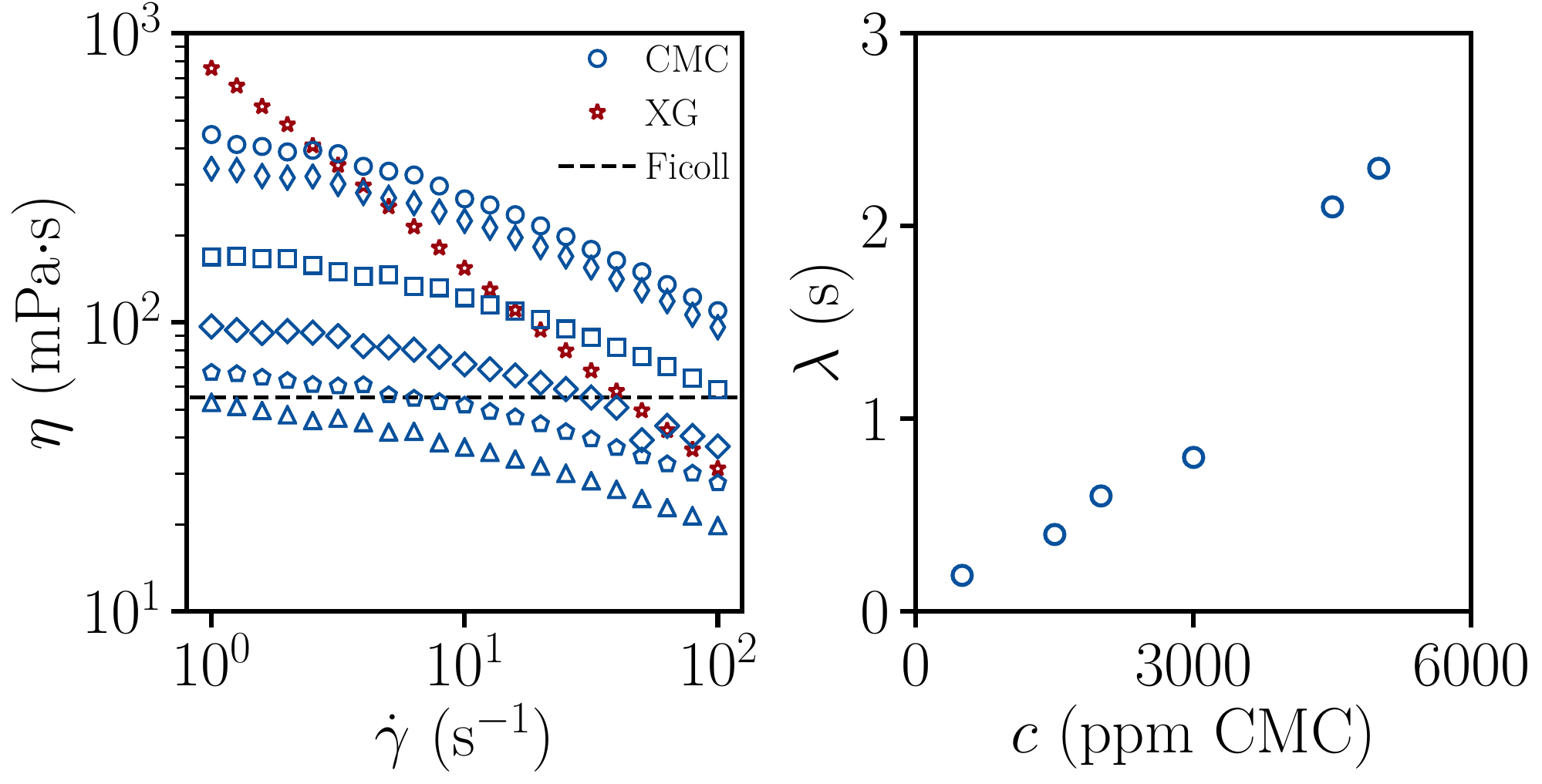}}
\caption{(Color available online) \textit{(a)} Steady shear viscosity $\eta$ as a function of shear rate $\dot{\gamma}$ for a Newtonian Ficoll suspension, viscoelastic CMC suspensions, where increasing viscosity indicates increasing concentration, and a shear-thinning XG suspension. \textit{(b)} Relaxation time $\lambda$ for viscoelastic CMC suspensions as a function of polymer concentration. }
\label{Rheology}
\end{figure}

Viscoelastic fluids are prepared by adding the biocompatible polymer sodium carboxymethyl cellulose (CMC) to deionized water (Sigma-Aldrich 419338, MW = $7 \times 10^5$ Da). CMC is a long flexible polymer that exhibits relaxation times on the order of seconds even at dilute concentrations ($c \ll c^* \approx 10000$ ppm, where $c^*$ is the overlap concentration). We characterized these suspensions using a cone-and-plate strain-controlled rheometer (TA Instruments RFS III). Figure~\ref{Rheology}\textit{(a)} shows the steady shear behavior of CMC suspensions~\cite{Shen2011}. Increasing CMC concentration yields increasing viscosity, from approximately 50 mPa$\cdot$s at 1000 ppm to 400 mPa$\cdot$s at 5000 ppm CMC. To control for the increased viscosity and shear-thinning effects of CMC, we examined two additional fluids: a Newtonian and a shear-thinning fluid. The Newtonian fluid is an aqueous Ficoll PM 400 solution with a concentration of 23\% by weight, which has a constant viscosity $\mu$ of approximately 55 mPa$\cdot$s (Sigma-Aldrich F4375). The shear-thinning fluids is a 3000 ppm solution of xanthan gum (XG, 2.7 $\times 10^6$ MW, Sigma Aldrich G1253) in M9 buffer solution; this fluid is strongly shear-thinning and has negligible elasticity compared to the CMC suspensions~\cite{Brenner1974, Shen2011, Gagnon2014, Gagnon2016, Gagnon2018}. Figure~\ref{Rheology}\textit{(b)} shows the dependence of the longest relaxation time on CMC concentration; increasing CMC concentration yields an increasing relaxation time from 0.18 s to 2.3 s between 500 and 5000 ppm~\cite{Patteson2015, Shen2011}. Relaxation times are obtained by measuring the fluid shear modulus $G(t)$ over time and fitting the data to the generalized linear elastic model of the form $G(t) = G_{0}e^{-t/\lambda{}}$~\cite{Patteson2015, Shen2011}.

\section{Results}

Unlike the unconfined case \cite{Shen2011}, we find a significant qualitative difference between the swimming gait of \textit{C. elegans} when confined in a Newtonian versus a viscoelastic fluid. Figure~\ref{Setup} shows the experimentally measured body shapes over one cycle for confined \textit{C. elegans} swimming through a \textit{(b)} a Newtonian and \textit{(c)} a viscoelastic fluid. The presence of viscoelasticity appears to substantially modify the \textit{C. elegans} beating patterns; in particular, the nematode's center-lines in the viscoelastic case show a markedly higher beating amplitude than the Newtonian case. To explore these changes, we systematically quantifying \textit{C. elegans} swimming kinematics in both Newtonian and viscoelastic fluids under confinement conditions ranging from unconfined to channels approximately equal to the length of the nematode. In order to quantify the effects of geometric confinement, we normalize the nematode's beating wavelength $\Lambda$ by the channel width $w$; values of $\Lambda/w \gtrsim 1$ indicate the swimmer will feel the effects of the nearby fluid-solid interfaces, while values of $\Lambda/w \ll 1$ indicate the swimmer is unconfined.

\subsection{Newtonian kinematics}

\begin{figure}[t]
\centerline{\includegraphics[width=0.5\textwidth]{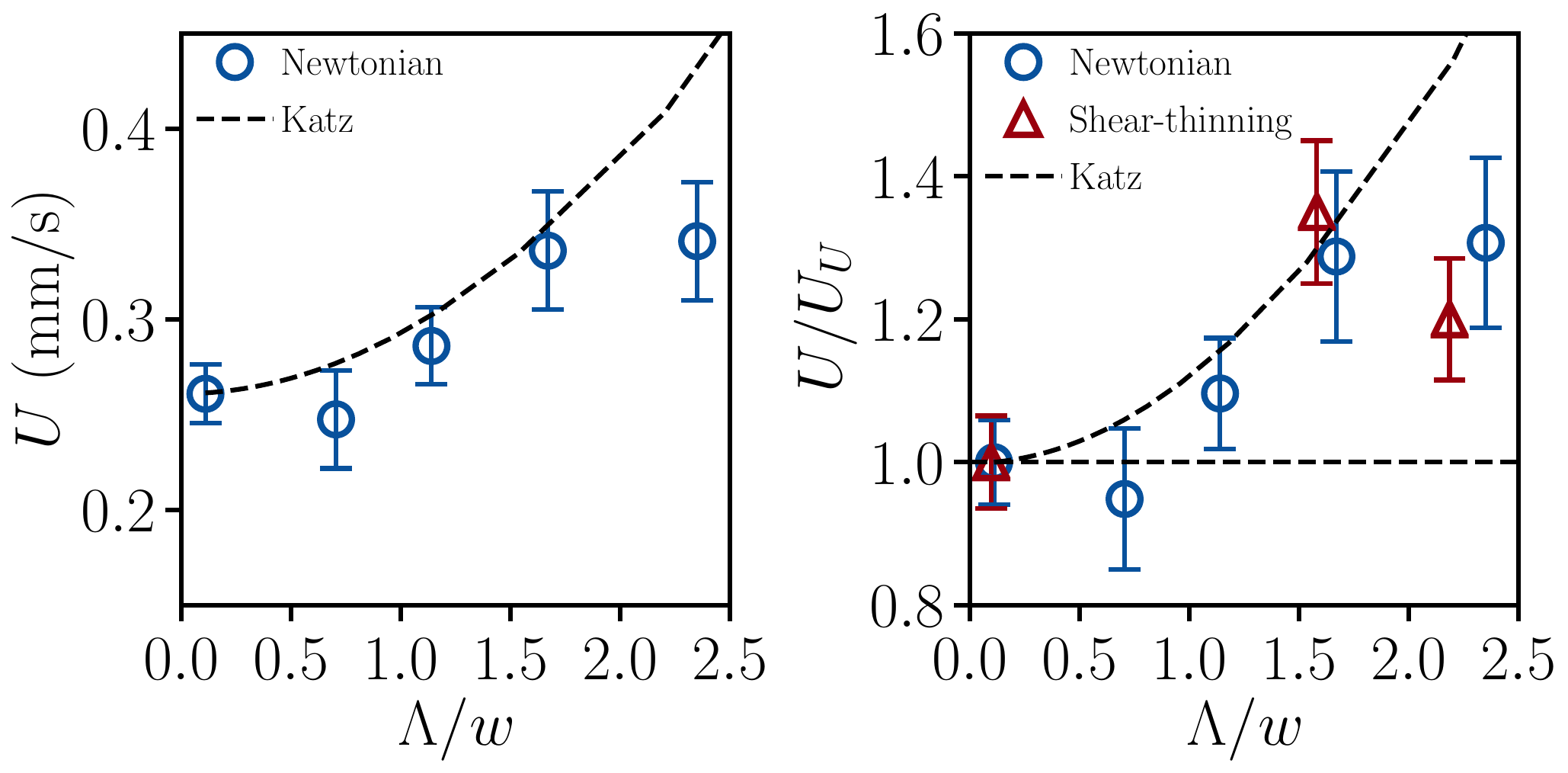}}
\caption{(Color available online) \textit{(a)} Swimming speed $U$ for Newtonian and shear-thinning fluids as a function of confinement $\Lambda/w$. \textit{(b)} Normalized speed $U/U_U$, showing that confinement results yields nearly a 40\% increase in swimming speed for both Newtonian and shear-thinning fluids.}
\label{GNewtSpeed}
\end{figure}

We begin by considering the effects of confinement on swimming speed in Newtonian fluids. Figure~\ref{GNewtSpeed}\textit{(a)} shows swimming speed versus confinement $\Lambda/w$ for five different geometries, ranging from effectively unconfined ($\Lambda/w \approx 0.1$) to confined ($\Lambda/w \gtrsim 1$). We observe an increase in swimming speed relative to the unconfined case for $\Lambda/w > 1$. We also experiment with a viscous generalized Newtonian (i.e., shear-thinning) fluid (XG 3000 ppm) to separate viscoelastic from shear-thinning effects.  We note that swimming speed also appears to increase with confinement, as perhaps expected from the Newtonian analysis \cite{Katz1974}. Figure~\ref{GNewtSpeed}\textit{(b)} shows the relative increase in swimming speed as a function of confinement compared to the  speed of unconfined \textit{C. elegans} in both Newtonian and generalized Newtonian fluids. Both cases suggest confined \textit{C. elegans} swim about 30\% faster when transversely confined. This increase in swimming speed for an undulatory gait in the presence of nearby boundaries is expected through analytical studies of Taylor's waving sheet~\cite{Katz1974}, numerical simulations of Taylor's waving cylinder~\cite{Felderhof2010}, and experiments with \textit{C. elegans} in structured environments~\cite{majmudar2012experiments}; this increase in speed is the result of increasing drag (propulsive) forces due to the presence of the walls. Nevertheless, our results establish a baseline for the viscoleastic case, which will be discussed below. 

\subsection{Viscoelastic kinematics}

\begin{figure}
\centerline{\includegraphics[width=.5\textwidth]{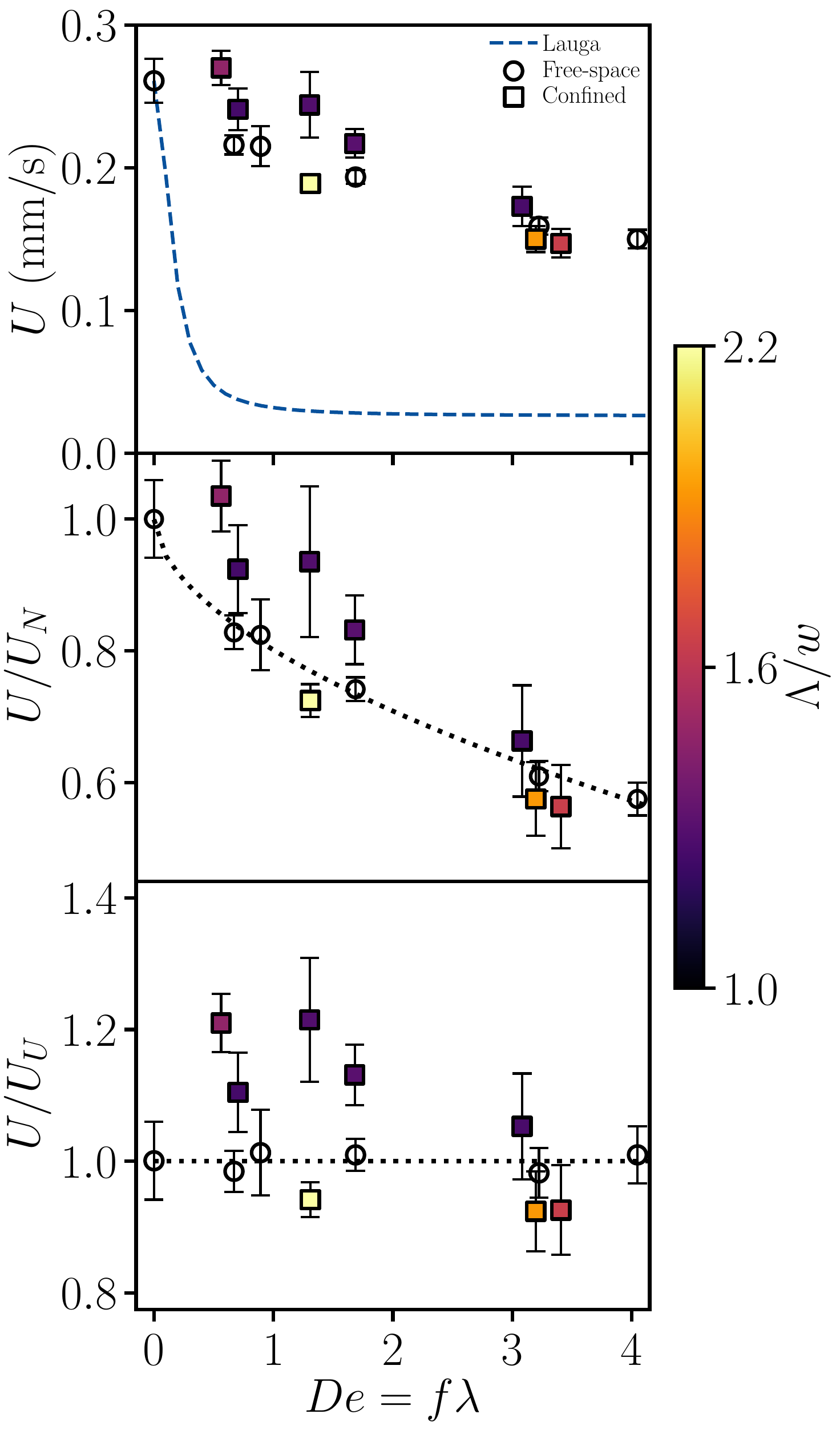}}
\caption{(Color available online)  \textit{(a)} Swimming speed as a function of $\De$ for all values of confinement. Line corresponds to fit with viscosity ratio of 0.1. \textit{(b)} Swimming speed in viscoelastic fluids normalized by the unconfined Newtonian swimming speed $U_N$. Color indicates the strength of confinement $\Lambda/w$. \textit{(c)} Swimming speed in viscoelastic fluids normalized by the unconfined speed $U_U$. Color indicates the strength of confinement.}
\label{VESpeed}
\end{figure}

Next, we consider confinement's effect on the nematode's swimming speed in viscoelastic fluids. Figure~\ref{VESpeed}\textit{(a)} shows \textit{C. elegans}' swimming speed as a function of Deborah number $De = f \lambda$. This non-dimensional quantity compares the characteristic timescale of \textit{C. elegans} beating gait, frequency $f$, and the longest relaxation time of the fluid $\lambda$ (see Fig~\ref{Rheology}\textit{(b)}). When $De \ll 1$, elastic stresses relax more quickly than the beating period and therefore minimally effect swimming kinematics. However, when $De \gtrapprox 1$, elastic stress are unable to relax fully within a period, and each new beating cycle occurs under the fading memory of the previous cycle. These remaining fluid stresses can then modify the subsequent kinematics of swimmer. For unconfined \textit{C. elegans}, we find that increasing viscoelasticity and therefore increasing $\De$ generally results in a diminished swimming speed compared to the Newtonian case, consistent with previous experimental~\cite{Shen2011} and numerical~\cite{Thomases2014} studies.

To more closely examine the effect of confinement, we now normalize our viscoelastic data by the unconfined Newtonian swimming speed $U_N$ (Fig.~\ref{VESpeed}\textit{(b)}. Furthermore, to guide the eye, we fit a polynomial to the unconfined normalized swimming speed $U/U_N$ (open circles), represented by the dashed line. We observe that five data points (each point corresponds to 10 experiments, N=10) lie above this line and three data points lie below. Color-coding by our previously identified non-dimensional confinement $\Lambda/w$, we observe that groups under moderate confinement ($\Lambda/w \approx 1$) lie above this unconfined speeds. Additionally, groups with strong confinement ($\Lambda/w \gtrapprox 1.5$) lie below the unconfined speeds. 

\begin{figure}
\centerline{\includegraphics[width=.5\textwidth]{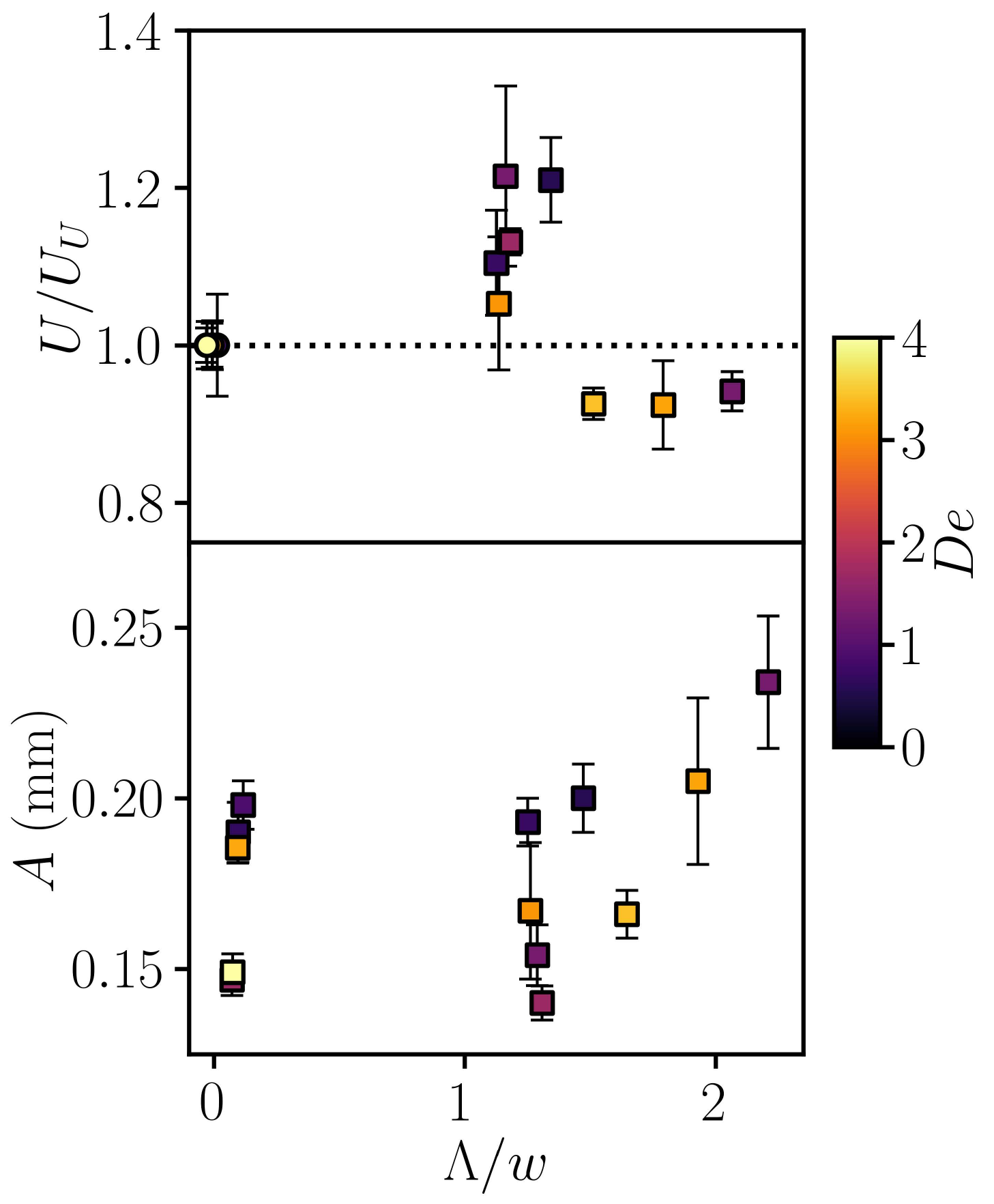}}
\caption{(Color available online)  \textit{(a)} Swimming speed in viscoelastic fluids normalized by the unconfined speed $U_U$ as a function of confinement strength $\Lambda/w$. Note the non-monotonic dependence of swimming speed on confinement is independent of $\De$ (color). \textit{(b)} Nematode amplitude measurements as a function of confinement strength $\Lambda/w.$ Color indicates $\De.$}
\label{speed_nonmonotonic}
\end{figure}

To examine the magnitude of this effect, we normalize speed $U$ by the unconfined speed $U_U$ at any given $\De$ (Fig.~\ref{VESpeed}\textit{(c)}). We can now more easily see that moderate confinement can increase swimming speed by as much as 20\%, while strong confinement decreases swimming speed by a little under 10\%; the effect of confinement depends nonlinearly on $\De$.  Another view of this data is shown in Figure~\ref{speed_nonmonotonic}\textit{(a)} where swimming speed is plotted as a function of confinement $\Lambda/w$, and color now indicates $\De$. The relationship between speed and confinement is clearly non-monotonic. Some confinement indeed helps \textit{C. elegans} swim faster, but unlike generalized Newtonian fluids, increasing confinement further yields speeds even slower than the unconfined case.

One possible explanation for the observed results may lie in the behavior of our swimmers. Thus, the next step is to examine the dependence of swimming gait on confinement more directly for viscoelastic fluids.  We qualitatively observed in Fig.~\ref{Setup}\textit{(b,c)} that viscoelastic fluids may increase the amplitude of \textit{C. elegans} under confinement. 
In Figure~\ref{speed_nonmonotonic}\textit{(b)} we plot the amplitude as a function of confinement with color indicating $\De.$  First we note that in a Newtonian fluid the amplitude does not change significantly with confinement, for those cases $A\approx 0.20.$ However with viscoelasticity  confinement can significantly change the amplitude of the swimmer gait. We find a reduction in amplitude by more than 25\% in weak confinement and an increase in amplitude by nearly 25\% in strong confinement. In addition in viscoelastic fluids the amplitude of the gait increases with increasing confinement.  For an unconfined swimmer in a Newtonian fluid, increasing amplitude would yield an increase in speed $U$~\cite{Taylor1951, Gray1955}, but this is not what is seen with \textit{C. elegans} in viscoelastic fluids.

\section{Swimming Analysis \& Discussion } 
Our experiments have shown that confinement in a Newtonian fluid increases \textit{C. elegans} swimming speed, as predicted by theoretical studies (Fig. \ref{GNewtSpeed}).  For viscoelastic fluids, on the other hand, we find that that weak or mild confinement leads to increases in swimming speed (relative to the unconfined swimmers), whereas stronger confinement leads to speed reductions (Fig. \ref{VESpeed}.  These changes in swimming speed cannot be attributed to changes in the Deborah number alone (Fig. \ref{speed_nonmonotonic}.  The gait of \textit{C. elegans} is also affected by confinement (Fig. \ref{Setup}, Fig. \ref{speed_nonmonotonic}b). While \textit{C. elegan} swimming amplitude is relatively insensitive to confinement in Newtonian fluids, we find that confinement in viscoelastic fluids appears to reduce the amplitude for weakly confined swimmers but increase the amplitude for highly confined swimmers. Although both increasing amplitude and increasing confinement speeds up \textit{C. elegans} propulsion in Newtonian fluids, we find the opposite effect in viscoelastic fluids, that is, reduction in swimming speed. Next, we explore the mechanisms for these observations.

Theoretical studies have investigated the effects of fluid elasticity for large amplitude, finite length undulatory swimmers in unconfined systems \cite{Teran2010, Thomases2014,thomases2017role, Thomases2019}.  Those studies indicate that elastic stresses that accumulate near swimmer bodies lead to swimming speed reductions.  While we cannot directly measure elastic stresses in our experiments, we argue here that measurable quantities in our velocimetry data (namely local strain rates) allow us to connect the experimentally observed swimming speed changes in viscoelastic fluids to the accumulation of large elastic stress near the swimmers.  Before addressing this issue, we explain how the previously developed theory applies to these experiments. 

As mentioned above, numerical simulations of undulatory swimming in viscoelastic fluids \cite{Teran2010,Thomases2014} have shown that large amplitude gaits lead to the accumulation of elastic stresses near (i.e, at the tip) of the swimmers. This stress accumulation leads to significant swimming speed reductions relative to swimmers in Newtonian fluids with the same large amplitude gaits. By contrast, low amplitude gaits do not lead to large stress accumulation, and relative speed differences are milder. Our experimental data (Figure~\ref{Setup}\textit{(b,c)}) show that, in a viscoelastic confinement environment, the swimmer amplitude is larger but the shape of the gait is similar.  Also, Fig.~\ref{Setup}\textit{(d,e)} shows that these larger amplitude gaits lead to faster fluid flow velocities and thus larger strain rates.  Recent analysis  \cite{Thomases2019} has shown that polymeric stresses accumulate near undulatory swimmers due to the development of an oscillatory extensional flow.  In that work, analytical and numerical simulations of a variety of oscillatory extensional flows, such as those that develop near bending filaments, are examined using the Oldroyd-B model for a viscoelastic fluid at zero Reynolds number. Significant nonlinear feedback is observed, leading to the development of large polymeric stresses for high-amplitude swimmers. In what follows, we will explore this idea by using a simple oscillatory extensional flow and our experimentally measured velocity fields.



We begin by considering a simple oscillating extensional flow of the form \begin{equation}\label{udef}\u(x,y,t)=\alpha h(t/T)(x,-y)\end{equation} where $h(t)$ is a periodic function with period $1$, mean zero, and maximum 1.   When $h(t)$ is a square wave it is possible to analyze this model problem in the Oldroyd-B \cite{Bird1987} system and to compute the maximum stress over a period. The non-dimensional stress is given by $\sig=\tauB/2\mu_p \alpha$, where $\tauB$ is the dimensional stress tensor, and $\mu_p$ is the polymer viscosity. The maximum in time of the stress (strain energy density) at the origin in this oscillating extensional flow can be computed analytically as 

\begin{figure}
\centerline{\includegraphics[width=.35\textwidth]{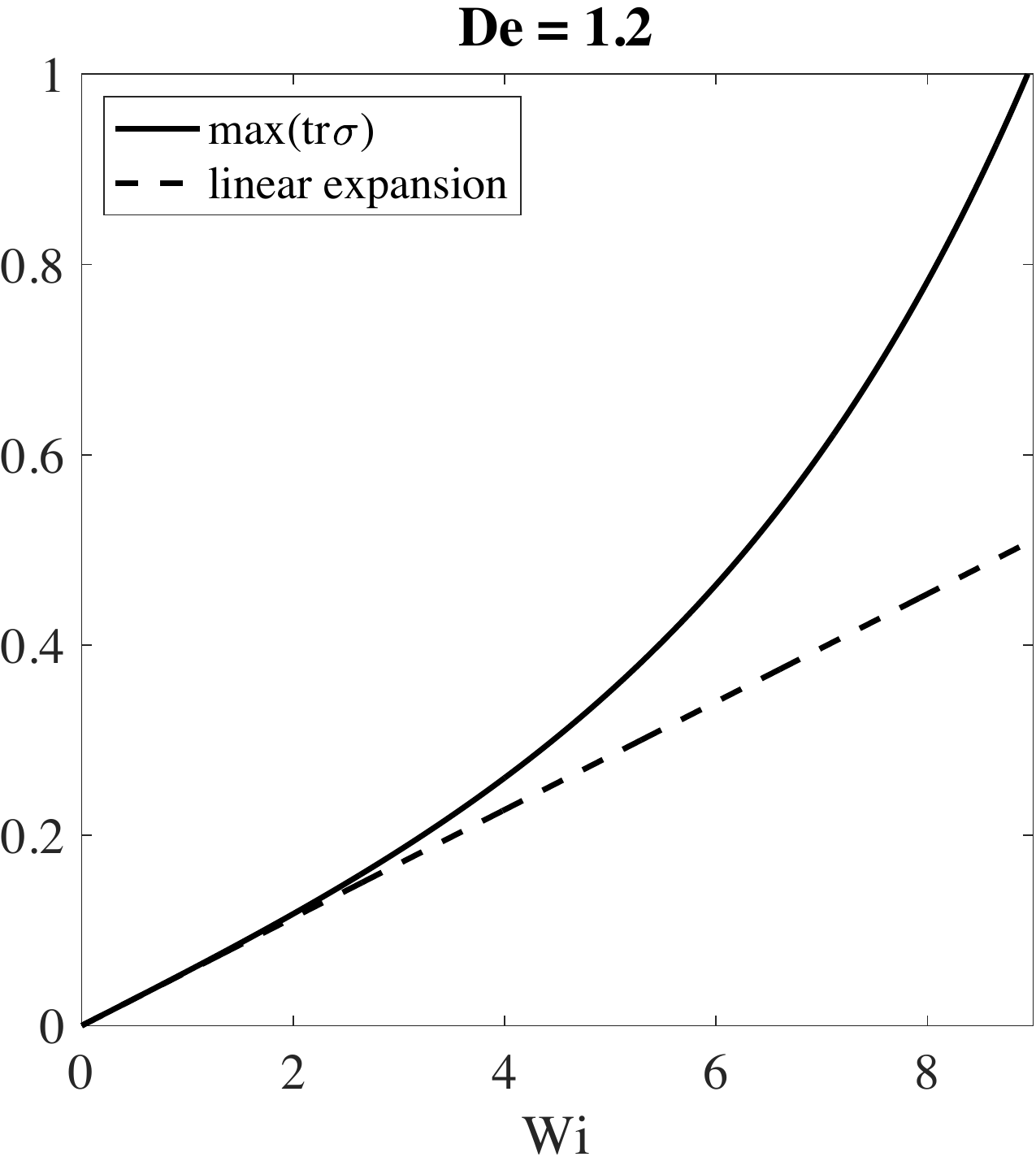}}
\caption{Plot of function given in Eq. \eqref{sig_theory} for $\De=1.2,$ along with the linear expansion of this function for small $\Wi.$ The functional form is obtained by assuming given oscillatory extensional flow in the Oldroyd-B model of a viscoelastic fluid. Note that 1.2 is chosen because this is the value that is used in experiments (c.f., Fig. \ref{WiFields})}
\label{mtrsigfig}
\end{figure} 

\begin{equation}\label{sig_theory}
   \max \tr\sig = 
   \frac{ 2\sinh\left(\frac{\Wi}{2\De}\right) 
         -2\Wi\sinh\left(\frac{1}{2\De}\right)}
        {(\Wi^2-1)\sinh\left(\frac{1}{2\De}\right)},
\end{equation}
where the Weissenberg number is $\Wi= 2\alpha\lambda,$ and
the Deborah number is $\De=\lambda/T$, for a given relaxation time $\lambda.$ We note that for this prescribed flow the strain rate is $\boldsymbol{\dot{\gamma}}\equiv\frac{1}{2} \left( \nabla \u + \nabla \u^{\mathrm{T}} \right)=2\alpha.$ and thus this definition of $\Wi$ agrees with the standard definition of $\Wi=\lambda\dot{\gamma}$.

Examining the functional form of $ \max \tr\sig$  we find two different regimes for
how the stress depends on $\Wi$. We also find a Deborah number-dependent
transition between the two regimes. To understand the behavior in the
two regimes, we expand the max trace of the stress in the limits of
large and small $\Wi$. For small $\Wi$, the max trace stress scales
linearly with $\Wi,$ while for large $\Wi$, the max trace of the stress to leading order is exponential in $\Wi.$ There is a smooth transition between these two regimes. We plot this function along with the linear expansion for small $\Wi$ at a fixed $\De=1.2$, as shown in Fig. \ref{mtrsigfig}. This value of $De$ is chosen to correspond with the experimental value of $De$ used in the experiments shown in Fig. \ref{Rheology}.  At this value of $\De$ we see that the stress response is linear for small $\Wi$ but the deviation from the linear expansion begins around $\Wi\approx 2$ where the fit differs from the functional form by about 3\% and by $\Wi\approx 4$ the deviation is closer to 15\%. This indicates that the stress response to the flow is in the linear regime for $\Wi\lesssim 2$ and in the nonlinear (or exponential) regime for $\Wi\gtrsim 2.$   

\begin{figure*}
\centerline{\includegraphics[width=\textwidth]{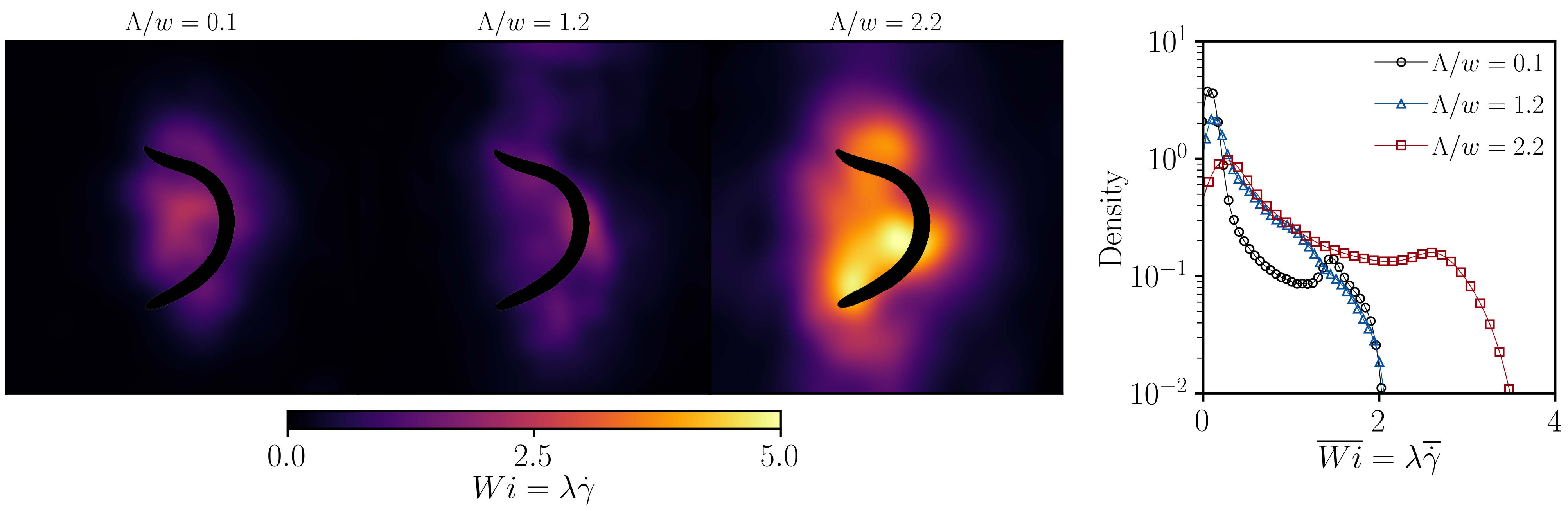}}
\caption{(Color available online) Spatial distribution of $\Wi = \lambda\dot{\gamma}$ for the phase when the \textit{C. elegans} is in the `C'-shaped configuration.  \textit{(left)} unconfined, \textit{(center)} weakly confined, and \textit{(right)} strongly confined \textit{C. elegans}. Right figure shows the density of the phase-averaged Weissenberg number as a function of confinement.}
\label{WiFields}
\end{figure*}


We note that the theory described above is applicable to the motion of undulatory swimmers by examining flows around oscillating filaments \cite{Thomases2019}. It is shown that the maximum in time of the strain rate near oscillating filaments scales linearly with the amplitude of the filament.  As in the case of fixed oscillating extensional flow, the anlaysis show that flows around oscillating filaments experience two different flow regimes with a smooth transition between these regimes (and this transition depends on $\De$) \cite{Thomases2019}; for low $\Wi$ (or amplitude) there is a linear stress response and for high $\Wi$ there is a nonlinear stress response leading to regions of highly concentrated polymer stress around the oscillating filaments. 

To apply this theory to our experimental data, we measure the velocity flow fields produced by the swimming nematodes. We focus our attention on three particular cases, all at $ De > 1$ ($De = 1.2$ ), (i) unconfined ($\Lambda/w = 0.1$, $U \approx U_U$), (ii) moderately confined ($\Lambda/w = 1.2$, $U > U_U$), and (iii) strongly confined ($\Lambda/w = 2.2$, $U<U_U$).  The instantaneous velocity fields are experimentally measured by adding 3.1~$\mu$m polystyrene tracer particles to the fluid medium; these particles are tracked continuously for the entire duration of the experiment using in-house tracking codes. These tracer particles are dilute ($<\!0.5$\% by volume) and do not alter the properties of the fluid. We image nematodes swimming for 6 to 10 beating cycles, where each cycle (or period) contains a minimum of 73 phases. We then phase-average the data to obtain highly  resolved and differentiable velocity fields. 

Samples of these velocity fields are shown in Fig.~\ref{Setup}\textit{(d,e)} for the unconfined and strongly confined cases, respectively. Color indicates fluid speed and instantaneous streamlines are overlaid in white. In this `C'-shaped configuration, the nematode's body is not rapidly moving, but the head and tail here are changing directions; in the preceding phases, the head and tail were moving closer together and the midsection of the nematode moved from left to right. Here, the head and tail have begun to move apart, and the midsection of the nematode is moving from right to left.  The velocities are significantly larger, with higher strain rates, for the confined swimmer. 

From these highly resolved velocity fields we compute the local strain rate tensor $\boldsymbol{\dot{\gamma}}=\frac{1}{2} \left( \nabla \u + \nabla \u^{\mathrm{T}} \right)$. We can fully resolve the full 3D shear rate tensor by applying a correction factor developed expressly to estimate the errors present in planar particle tracking data; without this, we would underestimate the shear rate magnitude and therefore the strength of viscoelastic effects by as much as 40\%~\cite{Montenegro2016}. Due to the predominately planar motion of \textit{C. elegans}, we are able to estimate out-of-plane shear rates through symmetry arguments and incompressibility. Assuming the nematode beats in the x-y plane, of the additional terms in the 3D versus 2D shear rate tensor, the only term that is non-zero is the z-gradient of fluid velocity in the z-direction $\mathrm{d} u_z/\mathrm{d}z$ (for more details, see~\cite{Montenegro2016} and~\cite{Gagnon2018}).

Figure~\ref{WiFields} (left panels) shows the spatial distribution of the local Weissenberg number, $\Wi = \lambda\boldsymbol{\dot{\gamma}}$ for the phase when the \textit{C. elegans} is in the `C'-shaped configuration,  \textit{(left)} unconfined, \textit{(center)} weakly confined, and \textit{(right)} strongly confined \textit{C. elegans}. While all three fields have regions of moderate $\Wi$ near the body, there is a clear and significant increase in $\Wi$  with increasing confinement.  For the snapshots displayed the unconfined and weakly confined nematodes are surrounded by regions of $\Wi\lesssim 2$ and the strongly confined nematode is surrounded by regions near the body of $\Wi\gtrsim 2.$  We can now infer stress accumulation around the nematodes using the the theory presented above. In Figure~\ref{mtrsigfig} we showed that for $\Wi\lesssim 2$ the flow is in a linear response regime for stress accumulation, while for $\Wi\gtrsim 2$ the flow is in a nonlinear response regime. 

This snapshot only shows a moment of time and hence in  Figure~\ref{WiFields} (right) we present the density of the phase averaged Weissenberg number $\overline{\Wi}=\lambda\overline{\dot{\gamma}}$ for the unconfined, weakly confined, and highly confined swimmers. We see that on average the unconfined and weakly confined nematodes experience flows with $\Wi<2$. The strongly confined nematode by contrast has a significant portion of the flow experiencing $\Wi>2$ on average over a cycle.  

We also point out the agreement between the densities for the two confined swimmers over $0.5\lesssim\overline{\Wi}\lesssim 1.5$. For higher values of $\overline{\Wi}$ these densities diverge and the unconfined and weakly confined agree for  $1.5\lesssim \overline{\Wi}\lesssim 2$.  This suggests that confinement increases $\Wi$. For some confinement there is a higher density of larger $\Wi$ values but it is only for strong confinement is the $\Wi$ increased into the nonlinear stress response regime. 

It has been observed that stress accumulation can slow down swimmers \cite{Thomases2019} and hence the connection with high $\Wi$ flows and stress response could explain how confinement effects the swimming speed in these experiments. We conjecture that for these nematodes confinement leads to faster swimming and larger amplitudes; together this leads to higher $\Wi$ flows and thus pushes the flow into a nonlinear elastic stress response regime where stress accumulation can lead to slow downs in swimming speed.  As was shown in \cite{Thomases2019} in regions of high elastic stress both $\De$ and $\Wi$ are needed to understand the effect of fluid elasticity on swimming. 

\section{Summary}
In this contribution, we analyze the motility behavior of an undulatory swimmer in viscoelastic fluids under confinement. Experiments are performed using the nematode \textit{C. elegans} in fluids with different levels of elasticity and geometries with different levels of confinement; experiments with viscous Newtonian fluids are also presented. Our results show that there is a complex interplay between fluid elasticity and confinement. That is, in viscoelastic fluids, the presence of boundaries can either increase or decrease the nematode's swimming speed. Whether speed is enhanced or diminished is dependent on the competing effects of beneficial viscous drag forces that can be enhanced by the presence of walls and detrimental elastic stresses. For viscoelastic fluids, weak and strong levels of confinement can lead to enhanced or hindered \textit{C. elegans} swimming speed, respectively. 

A recent analysis is used to understand the experimental observations by estimating the fluid elastic stresses in these swimming systems. Using experimentally measured velocity fields, we are able to directly measuring the growth of elastic stresses in the fluids.  We find that the observed hindered swimming speed at high level of confinement is due to higher $\Wi$flows; this pushes the flow into a nonlinear elastic stress response regime where stress accumulation can lead to slow downs in swimming speed.  Furthermore, there is a, $\De-$ dependent, $\Wi$  transition from a linear stress response regime to a nonlinear (or exponential) stress response regime.  Overall, our results show the importance of stretching history to locomotion under confinement. We speculate that this would also be case for bulk swimming in viscoelastic fluids.

\section*{acknowledgments}
We thank J. H. Shih and E. A. Fedalei for swimming movies, R. G. Kalb for \textit{C. elegans}, X. Shen, B. Qin, and A. E. Patteson for rheology data.

\bibliography{Ve_Conf_elegans.bib}

\begin{thebibliography}{64}%
\makeatletter
\providecommand \@ifxundefined [1]{%
 \@ifx{#1\undefined}
}%
\providecommand \@ifnum [1]{%
 \ifnum #1\expandafter \@firstoftwo
 \else \expandafter \@secondoftwo
 \fi
}%
\providecommand \@ifx [1]{%
 \ifx #1\expandafter \@firstoftwo
 \else \expandafter \@secondoftwo
 \fi
}%
\providecommand \natexlab [1]{#1}%
\providecommand \enquote  [1]{``#1''}%
\providecommand \bibnamefont  [1]{#1}%
\providecommand \bibfnamefont [1]{#1}%
\providecommand \citenamefont [1]{#1}%
\providecommand \href@noop [0]{\@secondoftwo}%
\providecommand \href [0]{\begingroup \@sanitize@url \@href}%
\providecommand \@href[1]{\@@startlink{#1}\@@href}%
\providecommand \@@href[1]{\endgroup#1\@@endlink}%
\providecommand \@sanitize@url [0]{\catcode `\\12\catcode `\$12\catcode `\&12\catcode `\#12\catcode `\^12\catcode `\_12\catcode `\%12\relax}%
\providecommand \@@startlink[1]{}%
\providecommand \@@endlink[0]{}%
\providecommand \url  [0]{\begingroup\@sanitize@url \@url }%
\providecommand \@url [1]{\endgroup\@href {#1}{\urlprefix }}%
\providecommand \urlprefix  [0]{URL }%
\providecommand \Eprint [0]{\href }%
\providecommand \doibase [0]{https://doi.org/}%
\providecommand \selectlanguage [0]{\@gobble}%
\providecommand \bibinfo  [0]{\@secondoftwo}%
\providecommand \bibfield  [0]{\@secondoftwo}%
\providecommand \translation [1]{[#1]}%
\providecommand \BibitemOpen [0]{}%
\providecommand \bibitemStop [0]{}%
\providecommand \bibitemNoStop [0]{.\EOS\space}%
\providecommand \EOS [0]{\spacefactor3000\relax}%
\providecommand \BibitemShut  [1]{\csname bibitem#1\endcsname}%
\let\auto@bib@innerbib\@empty
\bibitem [{\citenamefont {Lauga}\ and\ \citenamefont {Powers}(2009)}]{Lauga2009}%
  \BibitemOpen
  \bibfield  {author} {\bibinfo {author} {\bibfnamefont {E.}~\bibnamefont {Lauga}}\ and\ \bibinfo {author} {\bibfnamefont {T.}~\bibnamefont {Powers}},\ }\href@noop {} {\bibfield  {journal} {\bibinfo  {journal} {Reports on Progress in Physics}\ }\textbf {\bibinfo {volume} {72}},\ \bibinfo {pages} {096601} (\bibinfo {year} {2009})}\BibitemShut {NoStop}%
\bibitem [{\citenamefont {Spagnolie}(2015)}]{Spagnolie2015}%
  \BibitemOpen
  \bibinfo {editor} {\bibfnamefont {S.}~\bibnamefont {Spagnolie}},\ ed.,\ \href@noop {} {\emph {\bibinfo {title} {Complex Fluids in Biological Systems}}}\ (\bibinfo  {publisher} {Springer},\ \bibinfo {year} {2015})\BibitemShut {NoStop}%
\bibitem [{\citenamefont {Elfring}\ \emph {et~al.}(2015)\citenamefont {Elfring}, \citenamefont {Lauga},\ and\ \citenamefont {(ed.) Spagnolie}}]{Elfring2015}%
  \BibitemOpen
  \bibfield  {author} {\bibinfo {author} {\bibfnamefont {G.}~\bibnamefont {Elfring}}, \bibinfo {author} {\bibfnamefont {E.}~\bibnamefont {Lauga}},\ and\ \bibinfo {author} {\bibfnamefont {S.}~\bibnamefont {(ed.) Spagnolie}},\ }\href@noop {} {\emph {\bibinfo {title} {Complex Fluids in Biological Systems: Theory of Locomotion through Complex Fluids}}}\ (\bibinfo  {publisher} {Springer},\ \bibinfo {year} {2015})\BibitemShut {NoStop}%
\bibitem [{\citenamefont {Sznitman}\ \emph {et~al.}(2010{\natexlab{a}})\citenamefont {Sznitman}, \citenamefont {Shen}, \citenamefont {Sznitman},\ and\ \citenamefont {Arratia}}]{Sznitman2010PoF}%
  \BibitemOpen
  \bibfield  {author} {\bibinfo {author} {\bibfnamefont {J.}~\bibnamefont {Sznitman}}, \bibinfo {author} {\bibfnamefont {X.}~\bibnamefont {Shen}}, \bibinfo {author} {\bibfnamefont {R.}~\bibnamefont {Sznitman}},\ and\ \bibinfo {author} {\bibfnamefont {P.}~\bibnamefont {Arratia}},\ }\href@noop {} {\bibfield  {journal} {\bibinfo  {journal} {Phys. Fluids}\ }\textbf {\bibinfo {volume} {22}},\ \bibinfo {pages} {121901} (\bibinfo {year} {2010}{\natexlab{a}})}\BibitemShut {NoStop}%
\bibitem [{\citenamefont {Taylor}(1951)}]{Taylor1951}%
  \BibitemOpen
  \bibfield  {author} {\bibinfo {author} {\bibfnamefont {G.}~\bibnamefont {Taylor}},\ }\href@noop {} {\bibfield  {journal} {\bibinfo  {journal} {Proc. R. Soc. Lon. Ser.-A}\ }\textbf {\bibinfo {volume} {209}},\ \bibinfo {pages} {447} (\bibinfo {year} {1951})}\BibitemShut {NoStop}%
\bibitem [{\citenamefont {Lighthill}(1976)}]{Lighthill1976}%
  \BibitemOpen
  \bibfield  {author} {\bibinfo {author} {\bibfnamefont {J.}~\bibnamefont {Lighthill}},\ }\href@noop {} {\bibfield  {journal} {\bibinfo  {journal} {SIAM Rev.}\ }\textbf {\bibinfo {volume} {18}},\ \bibinfo {pages} {161} (\bibinfo {year} {1976})}\BibitemShut {NoStop}%
\bibitem [{\citenamefont {Korta}\ \emph {et~al.}(2007)\citenamefont {Korta}, \citenamefont {Clark},\ and\ \citenamefont {Gabel}}]{Korta2007}%
  \BibitemOpen
  \bibfield  {author} {\bibinfo {author} {\bibfnamefont {J.}~\bibnamefont {Korta}}, \bibinfo {author} {\bibfnamefont {D.}~\bibnamefont {Clark}},\ and\ \bibinfo {author} {\bibfnamefont {C.}~\bibnamefont {Gabel}},\ }\href@noop {} {\bibfield  {journal} {\bibinfo  {journal} {J. of Exp. Biol.}\ }\textbf {\bibinfo {volume} {210}},\ \bibinfo {pages} {2383} (\bibinfo {year} {2007})}\BibitemShut {NoStop}%
\bibitem [{\citenamefont {Guasto}\ \emph {et~al.}(2010)\citenamefont {Guasto}, \citenamefont {Johnson},\ and\ \citenamefont {Gollub}}]{Guasto2010}%
  \BibitemOpen
  \bibfield  {author} {\bibinfo {author} {\bibfnamefont {J.}~\bibnamefont {Guasto}}, \bibinfo {author} {\bibfnamefont {K.}~\bibnamefont {Johnson}},\ and\ \bibinfo {author} {\bibfnamefont {J.}~\bibnamefont {Gollub}},\ }\href@noop {} {\bibfield  {journal} {\bibinfo  {journal} {Phys. Rev. Lett.}\ }\textbf {\bibinfo {volume} {105}},\ \bibinfo {pages} {168102} (\bibinfo {year} {2010})}\BibitemShut {NoStop}%
\bibitem [{\citenamefont {Padmanabhan}\ \emph {et~al.}(2012)\citenamefont {Padmanabhan}, \citenamefont {Khan}, \citenamefont {Solomon}, \citenamefont {Armstrong}, \citenamefont {Rumbaugh}, \citenamefont {Vanapalli},\ and\ \citenamefont {Blawzdziewicz}}]{Padmanabhan2012}%
  \BibitemOpen
  \bibfield  {author} {\bibinfo {author} {\bibfnamefont {V.}~\bibnamefont {Padmanabhan}}, \bibinfo {author} {\bibfnamefont {Z.}~\bibnamefont {Khan}}, \bibinfo {author} {\bibfnamefont {D.}~\bibnamefont {Solomon}}, \bibinfo {author} {\bibfnamefont {A.}~\bibnamefont {Armstrong}}, \bibinfo {author} {\bibfnamefont {K.}~\bibnamefont {Rumbaugh}}, \bibinfo {author} {\bibfnamefont {S.}~\bibnamefont {Vanapalli}},\ and\ \bibinfo {author} {\bibfnamefont {J.}~\bibnamefont {Blawzdziewicz}},\ }\href@noop {} {\bibfield  {journal} {\bibinfo  {journal} {PloS One}\ }\textbf {\bibinfo {volume} {7}},\ \bibinfo {pages} {e40121} (\bibinfo {year} {2012})}\BibitemShut {NoStop}%
\bibitem [{\citenamefont {Bilbao}\ \emph {et~al.}(2013)\citenamefont {Bilbao}, \citenamefont {Wajnryb}, \citenamefont {Vanapalli},\ and\ \citenamefont {Blawzdziewicz}}]{Bilbao2013}%
  \BibitemOpen
  \bibfield  {author} {\bibinfo {author} {\bibfnamefont {A.}~\bibnamefont {Bilbao}}, \bibinfo {author} {\bibfnamefont {E.}~\bibnamefont {Wajnryb}}, \bibinfo {author} {\bibfnamefont {S.}~\bibnamefont {Vanapalli}},\ and\ \bibinfo {author} {\bibfnamefont {J.}~\bibnamefont {Blawzdziewicz}},\ }\href@noop {} {\bibfield  {journal} {\bibinfo  {journal} {Phys. Fluids}\ }\textbf {\bibinfo {volume} {25}},\ \bibinfo {pages} {081902} (\bibinfo {year} {2013})}\BibitemShut {NoStop}%
\bibitem [{\citenamefont {Spagnolie}\ and\ \citenamefont {Underhill}(2023)}]{Spagnolie_ARFM_2023}%
  \BibitemOpen
  \bibfield  {author} {\bibinfo {author} {\bibfnamefont {S.~E.}\ \bibnamefont {Spagnolie}}\ and\ \bibinfo {author} {\bibfnamefont {P.~T.}\ \bibnamefont {Underhill}},\ }\href {https://doi.org/10.1146/annurev-conmatphys-040821-112149} {\bibfield  {journal} {\bibinfo  {journal} {Ann. Rev. Cond. Matt. Phys.}\ }\textbf {\bibinfo {volume} {14}},\ \bibinfo {pages} {381} (\bibinfo {year} {2023})}\BibitemShut {NoStop}%
\bibitem [{\citenamefont {Arratia}(2022)}]{Arratia_PRF2022_Complex}%
  \BibitemOpen
  \bibfield  {author} {\bibinfo {author} {\bibfnamefont {P.~E.}\ \bibnamefont {Arratia}},\ }\href {https://doi.org/10.1103/PhysRevFluids.7.110515} {\bibfield  {journal} {\bibinfo  {journal} {Phys. Rev. Fluids}\ }\textbf {\bibinfo {volume} {7}},\ \bibinfo {pages} {110515} (\bibinfo {year} {2022})}\BibitemShut {NoStop}%
\bibitem [{\citenamefont {Fauci}\ and\ \citenamefont {Dillon}(2006)}]{Fauci2006}%
  \BibitemOpen
  \bibfield  {author} {\bibinfo {author} {\bibfnamefont {L.}~\bibnamefont {Fauci}}\ and\ \bibinfo {author} {\bibfnamefont {R.}~\bibnamefont {Dillon}},\ }\href@noop {} {\bibfield  {journal} {\bibinfo  {journal} {Annu. Rev. Fluid Mech.}\ }\textbf {\bibinfo {volume} {38}},\ \bibinfo {pages} {371} (\bibinfo {year} {2006})}\BibitemShut {NoStop}%
\bibitem [{\citenamefont {Lauga}(2007)}]{Lauga2007}%
  \BibitemOpen
  \bibfield  {author} {\bibinfo {author} {\bibfnamefont {E.}~\bibnamefont {Lauga}},\ }\href@noop {} {\bibfield  {journal} {\bibinfo  {journal} {Phys. Fluids}\ }\textbf {\bibinfo {volume} {19}},\ \bibinfo {pages} {083104} (\bibinfo {year} {2007})}\BibitemShut {NoStop}%
\bibitem [{\citenamefont {Patteson}\ \emph {et~al.}(2016)\citenamefont {Patteson}, \citenamefont {Gopinath},\ and\ \citenamefont {Arratia}}]{Patteson2016}%
  \BibitemOpen
  \bibfield  {author} {\bibinfo {author} {\bibfnamefont {A.~E.}\ \bibnamefont {Patteson}}, \bibinfo {author} {\bibfnamefont {A.}~\bibnamefont {Gopinath}},\ and\ \bibinfo {author} {\bibfnamefont {P.~E.}\ \bibnamefont {Arratia}},\ }\href@noop {} {\bibfield  {journal} {\bibinfo  {journal} {Current Opinion in Colloid \& Interface Science}\ }\textbf {\bibinfo {volume} {21}},\ \bibinfo {pages} {86} (\bibinfo {year} {2016})}\BibitemShut {NoStop}%
\bibitem [{\citenamefont {Shen}\ and\ \citenamefont {Arratia}(2011)}]{Shen2011}%
  \BibitemOpen
  \bibfield  {author} {\bibinfo {author} {\bibfnamefont {X.}~\bibnamefont {Shen}}\ and\ \bibinfo {author} {\bibfnamefont {P.}~\bibnamefont {Arratia}},\ }\href@noop {} {\bibfield  {journal} {\bibinfo  {journal} {Phys. Rev. Lett.}\ }\textbf {\bibinfo {volume} {106}},\ \bibinfo {pages} {208101} (\bibinfo {year} {2011})}\BibitemShut {NoStop}%
\bibitem [{\citenamefont {Larson}(1999)}]{Larson1999}%
  \BibitemOpen
  \bibfield  {author} {\bibinfo {author} {\bibfnamefont {R.}~\bibnamefont {Larson}},\ }\href@noop {} {\emph {\bibinfo {title} {The structure and rheology of complex fluids}}}\ (\bibinfo  {publisher} {Oxford University Press},\ \bibinfo {address} {New York},\ \bibinfo {year} {1999})\BibitemShut {NoStop}%
\bibitem [{\citenamefont {Fu}\ \emph {et~al.}(2009)\citenamefont {Fu}, \citenamefont {Wolgemuth},\ and\ \citenamefont {Powers}}]{Fu2009}%
  \BibitemOpen
  \bibfield  {author} {\bibinfo {author} {\bibfnamefont {H.}~\bibnamefont {Fu}}, \bibinfo {author} {\bibfnamefont {C.}~\bibnamefont {Wolgemuth}},\ and\ \bibinfo {author} {\bibfnamefont {T.}~\bibnamefont {Powers}},\ }\href@noop {} {\bibfield  {journal} {\bibinfo  {journal} {Phys. Fluids}\ }\textbf {\bibinfo {volume} {21}},\ \bibinfo {pages} {033102} (\bibinfo {year} {2009})}\BibitemShut {NoStop}%
\bibitem [{\citenamefont {Leshansky}(2009)}]{Leshansky2009}%
  \BibitemOpen
  \bibfield  {author} {\bibinfo {author} {\bibfnamefont {A.}~\bibnamefont {Leshansky}},\ }\href@noop {} {\bibfield  {journal} {\bibinfo  {journal} {Phys. Rev. E}\ }\textbf {\bibinfo {volume} {80}},\ \bibinfo {pages} {051911} (\bibinfo {year} {2009})}\BibitemShut {NoStop}%
\bibitem [{\citenamefont {Fu}\ \emph {et~al.}(2010)\citenamefont {Fu}, \citenamefont {Shenoy},\ and\ \citenamefont {Powers}}]{Fu2010}%
  \BibitemOpen
  \bibfield  {author} {\bibinfo {author} {\bibfnamefont {H.}~\bibnamefont {Fu}}, \bibinfo {author} {\bibfnamefont {V.}~\bibnamefont {Shenoy}},\ and\ \bibinfo {author} {\bibfnamefont {T.}~\bibnamefont {Powers}},\ }\href@noop {} {\bibfield  {journal} {\bibinfo  {journal} {Europhys. Lett.}\ }\textbf {\bibinfo {volume} {91}} (\bibinfo {year} {2010})}\BibitemShut {NoStop}%
\bibitem [{\citenamefont {Teran}\ \emph {et~al.}(2010)\citenamefont {Teran}, \citenamefont {Fauci},\ and\ \citenamefont {Shelley}}]{Teran2010}%
  \BibitemOpen
  \bibfield  {author} {\bibinfo {author} {\bibfnamefont {J.}~\bibnamefont {Teran}}, \bibinfo {author} {\bibfnamefont {L.}~\bibnamefont {Fauci}},\ and\ \bibinfo {author} {\bibfnamefont {M.}~\bibnamefont {Shelley}},\ }\href@noop {} {\bibfield  {journal} {\bibinfo  {journal} {Phys. Rev. Lett.}\ }\textbf {\bibinfo {volume} {104}},\ \bibinfo {pages} {038101} (\bibinfo {year} {2010})}\BibitemShut {NoStop}%
\bibitem [{\citenamefont {Juarez}\ \emph {et~al.}(2010)\citenamefont {Juarez}, \citenamefont {Lu}, \citenamefont {Sznitman},\ and\ \citenamefont {Arratia}}]{Juarez2010}%
  \BibitemOpen
  \bibfield  {author} {\bibinfo {author} {\bibfnamefont {G.}~\bibnamefont {Juarez}}, \bibinfo {author} {\bibfnamefont {K.}~\bibnamefont {Lu}}, \bibinfo {author} {\bibfnamefont {J.}~\bibnamefont {Sznitman}},\ and\ \bibinfo {author} {\bibfnamefont {P.~E.}\ \bibnamefont {Arratia}},\ }\href@noop {} {\bibfield  {journal} {\bibinfo  {journal} {Europhys. Lett.}\ }\textbf {\bibinfo {volume} {92}},\ \bibinfo {pages} {44002} (\bibinfo {year} {2010})}\BibitemShut {NoStop}%
\bibitem [{\citenamefont {Liu}\ \emph {et~al.}(2011)\citenamefont {Liu}, \citenamefont {Powers},\ and\ \citenamefont {Breuer}}]{Liu2011}%
  \BibitemOpen
  \bibfield  {author} {\bibinfo {author} {\bibfnamefont {B.}~\bibnamefont {Liu}}, \bibinfo {author} {\bibfnamefont {T.}~\bibnamefont {Powers}},\ and\ \bibinfo {author} {\bibfnamefont {K.}~\bibnamefont {Breuer}},\ }\href@noop {} {\bibfield  {journal} {\bibinfo  {journal} {Proc. Natl. Acad. Sci. USA}\ }\textbf {\bibinfo {volume} {108}},\ \bibinfo {pages} {19516} (\bibinfo {year} {2011})}\BibitemShut {NoStop}%
\bibitem [{\citenamefont {Harman}\ \emph {et~al.}(2012)\citenamefont {Harman}, \citenamefont {Dunham-Ems}, \citenamefont {Caimano}, \citenamefont {Belperron}, \citenamefont {Bockenstedt}, \citenamefont {Fu}, \citenamefont {Radolf},\ and\ \citenamefont {Wolgemuth}}]{Harman2012}%
  \BibitemOpen
  \bibfield  {author} {\bibinfo {author} {\bibfnamefont {M.}~\bibnamefont {Harman}}, \bibinfo {author} {\bibfnamefont {S.}~\bibnamefont {Dunham-Ems}}, \bibinfo {author} {\bibfnamefont {M.}~\bibnamefont {Caimano}}, \bibinfo {author} {\bibfnamefont {A.}~\bibnamefont {Belperron}}, \bibinfo {author} {\bibfnamefont {L.}~\bibnamefont {Bockenstedt}}, \bibinfo {author} {\bibfnamefont {H.}~\bibnamefont {Fu}}, \bibinfo {author} {\bibfnamefont {J.}~\bibnamefont {Radolf}},\ and\ \bibinfo {author} {\bibfnamefont {C.}~\bibnamefont {Wolgemuth}},\ }\href@noop {} {\bibfield  {journal} {\bibinfo  {journal} {Proc. Natl. Acad. Sci. USA}\ }\textbf {\bibinfo {volume} {109}},\ \bibinfo {pages} {3059} (\bibinfo {year} {2012})}\BibitemShut {NoStop}%
\bibitem [{\citenamefont {Gagnon}\ \emph {et~al.}(2013)\citenamefont {Gagnon}, \citenamefont {Shen},\ and\ \citenamefont {Arratia}}]{Gagnon2013}%
  \BibitemOpen
  \bibfield  {author} {\bibinfo {author} {\bibfnamefont {D.}~\bibnamefont {Gagnon}}, \bibinfo {author} {\bibfnamefont {X.}~\bibnamefont {Shen}},\ and\ \bibinfo {author} {\bibfnamefont {P.}~\bibnamefont {Arratia}},\ }\href@noop {} {\bibfield  {journal} {\bibinfo  {journal} {Europhys. Lett.}\ }\textbf {\bibinfo {volume} {104}},\ \bibinfo {pages} {14004} (\bibinfo {year} {2013})}\BibitemShut {NoStop}%
\bibitem [{\citenamefont {Gagnon}\ \emph {et~al.}(2014{\natexlab{a}})\citenamefont {Gagnon}, \citenamefont {Keim}, \citenamefont {Shen},\ and\ \citenamefont {Arratia}}]{Gagnon2014FIP}%
  \BibitemOpen
  \bibfield  {author} {\bibinfo {author} {\bibfnamefont {D.}~\bibnamefont {Gagnon}}, \bibinfo {author} {\bibfnamefont {N.}~\bibnamefont {Keim}}, \bibinfo {author} {\bibfnamefont {X.}~\bibnamefont {Shen}},\ and\ \bibinfo {author} {\bibfnamefont {P.}~\bibnamefont {Arratia}},\ }\href@noop {} {\bibfield  {journal} {\bibinfo  {journal} {Phys. Fluids}\ }\textbf {\bibinfo {volume} {26}},\ \bibinfo {pages} {103101} (\bibinfo {year} {2014}{\natexlab{a}})}\BibitemShut {NoStop}%
\bibitem [{\citenamefont {Patteson}\ \emph {et~al.}(2015)\citenamefont {Patteson}, \citenamefont {Gopinath}, \citenamefont {Goulian},\ and\ \citenamefont {Arratia}}]{Patteson2015}%
  \BibitemOpen
  \bibfield  {author} {\bibinfo {author} {\bibfnamefont {A.}~\bibnamefont {Patteson}}, \bibinfo {author} {\bibfnamefont {A.}~\bibnamefont {Gopinath}}, \bibinfo {author} {\bibfnamefont {M.}~\bibnamefont {Goulian}},\ and\ \bibinfo {author} {\bibfnamefont {P.}~\bibnamefont {Arratia}},\ }\href@noop {} {\bibfield  {journal} {\bibinfo  {journal} {Sci. Rep.}\ }\textbf {\bibinfo {volume} {5}},\ \bibinfo {pages} {15761} (\bibinfo {year} {2015})}\BibitemShut {NoStop}%
\bibitem [{\citenamefont {Qin}\ \emph {et~al.}(2015)\citenamefont {Qin}, \citenamefont {Gopinath}, \citenamefont {Yang}, \citenamefont {Gollub},\ and\ \citenamefont {Arratia}}]{Qin2015}%
  \BibitemOpen
  \bibfield  {author} {\bibinfo {author} {\bibfnamefont {B.}~\bibnamefont {Qin}}, \bibinfo {author} {\bibfnamefont {A.}~\bibnamefont {Gopinath}}, \bibinfo {author} {\bibfnamefont {J.}~\bibnamefont {Yang}}, \bibinfo {author} {\bibfnamefont {J.}~\bibnamefont {Gollub}},\ and\ \bibinfo {author} {\bibfnamefont {P.}~\bibnamefont {Arratia}},\ }\href@noop {} {\bibfield  {journal} {\bibinfo  {journal} {Sci. Rep.}\ }\textbf {\bibinfo {volume} {5}},\ \bibinfo {pages} {9190} (\bibinfo {year} {2015})}\BibitemShut {NoStop}%
\bibitem [{\citenamefont {Martinez}\ \emph {et~al.}(2014)\citenamefont {Martinez}, \citenamefont {Schwarz-Linek}, \citenamefont {Reufer}, \citenamefont {Wilson}, \citenamefont {Morozov},\ and\ \citenamefont {Poon}}]{martinez_pnas_2014}%
  \BibitemOpen
  \bibfield  {author} {\bibinfo {author} {\bibfnamefont {V.~A.}\ \bibnamefont {Martinez}}, \bibinfo {author} {\bibfnamefont {J.}~\bibnamefont {Schwarz-Linek}}, \bibinfo {author} {\bibfnamefont {M.}~\bibnamefont {Reufer}}, \bibinfo {author} {\bibfnamefont {L.~G.}\ \bibnamefont {Wilson}}, \bibinfo {author} {\bibfnamefont {A.~N.}\ \bibnamefont {Morozov}},\ and\ \bibinfo {author} {\bibfnamefont {W.~C.~K.}\ \bibnamefont {Poon}},\ }\href {https://doi.org/10.1073/pnas.1415460111} {\bibfield  {journal} {\bibinfo  {journal} {Proceedings of the National Academy of Sciences}\ }\textbf {\bibinfo {volume} {111}},\ \bibinfo {pages} {17771} (\bibinfo {year} {2014})}\BibitemShut {NoStop}%
\bibitem [{\citenamefont {Thomases}\ and\ \citenamefont {Guy}(2014)}]{Thomases2014}%
  \BibitemOpen
  \bibfield  {author} {\bibinfo {author} {\bibfnamefont {B.}~\bibnamefont {Thomases}}\ and\ \bibinfo {author} {\bibfnamefont {R.}~\bibnamefont {Guy}},\ }\href@noop {} {\bibfield  {journal} {\bibinfo  {journal} {Phys. Rev. Lett.}\ }\textbf {\bibinfo {volume} {113}},\ \bibinfo {pages} {098102} (\bibinfo {year} {2014})}\BibitemShut {NoStop}%
\bibitem [{\citenamefont {Thomases}\ and\ \citenamefont {Guy}(2019)}]{Thomases2019}%
  \BibitemOpen
  \bibfield  {author} {\bibinfo {author} {\bibfnamefont {B.}~\bibnamefont {Thomases}}\ and\ \bibinfo {author} {\bibfnamefont {R.~D.}\ \bibnamefont {Guy}},\ }\href@noop {} {\bibfield  {journal} {\bibinfo  {journal} {J. Non-Newt. Fluid Mech.}\ }\textbf {\bibinfo {volume} {269}},\ \bibinfo {pages} {47} (\bibinfo {year} {2019})}\BibitemShut {NoStop}%
\bibitem [{\citenamefont {Riley}\ and\ \citenamefont {Lauga}(2014)}]{Riley_2014}%
  \BibitemOpen
  \bibfield  {author} {\bibinfo {author} {\bibfnamefont {E.~E.}\ \bibnamefont {Riley}}\ and\ \bibinfo {author} {\bibfnamefont {E.}~\bibnamefont {Lauga}},\ }\href {https://doi.org/10.1209/0295-5075/108/34003} {\bibfield  {journal} {\bibinfo  {journal} {Europhys. Lett.}\ }\textbf {\bibinfo {volume} {108}},\ \bibinfo {pages} {34003} (\bibinfo {year} {2014})}\BibitemShut {NoStop}%
\bibitem [{\citenamefont {Sznitman}\ \emph {et~al.}(2010{\natexlab{b}})\citenamefont {Sznitman}, \citenamefont {Purohit}, \citenamefont {Krajacic}, \citenamefont {Lamitina},\ and\ \citenamefont {Arratia}}]{Sznitman2010BJ}%
  \BibitemOpen
  \bibfield  {author} {\bibinfo {author} {\bibfnamefont {J.}~\bibnamefont {Sznitman}}, \bibinfo {author} {\bibfnamefont {P.}~\bibnamefont {Purohit}}, \bibinfo {author} {\bibfnamefont {P.}~\bibnamefont {Krajacic}}, \bibinfo {author} {\bibfnamefont {T.}~\bibnamefont {Lamitina}},\ and\ \bibinfo {author} {\bibfnamefont {P.}~\bibnamefont {Arratia}},\ }\href@noop {} {\bibfield  {journal} {\bibinfo  {journal} {Biophys. J.}\ }\textbf {\bibinfo {volume} {98}},\ \bibinfo {pages} {617} (\bibinfo {year} {2010}{\natexlab{b}})}\BibitemShut {NoStop}%
\bibitem [{\citenamefont {Lauga}\ \emph {et~al.}(2006)\citenamefont {Lauga}, \citenamefont {DiLuzio}, \citenamefont {Whitesides},\ and\ \citenamefont {Stone}}]{Stone_2006}%
  \BibitemOpen
  \bibfield  {author} {\bibinfo {author} {\bibfnamefont {E.}~\bibnamefont {Lauga}}, \bibinfo {author} {\bibfnamefont {W.~R.}\ \bibnamefont {DiLuzio}}, \bibinfo {author} {\bibfnamefont {G.~M.}\ \bibnamefont {Whitesides}},\ and\ \bibinfo {author} {\bibfnamefont {H.~A.}\ \bibnamefont {Stone}},\ }\href {https://doi.org/https://doi.org/10.1529/biophysj.105.069401} {\bibfield  {journal} {\bibinfo  {journal} {Biophys. J.}\ }\textbf {\bibinfo {volume} {90}},\ \bibinfo {pages} {400} (\bibinfo {year} {2006})}\BibitemShut {NoStop}%
\bibitem [{\citenamefont {Turci}\ and\ \citenamefont {Wilding}(2021)}]{Wilding2021_PRL}%
  \BibitemOpen
  \bibfield  {author} {\bibinfo {author} {\bibfnamefont {F.}~\bibnamefont {Turci}}\ and\ \bibinfo {author} {\bibfnamefont {N.~B.}\ \bibnamefont {Wilding}},\ }\href {https://doi.org/10.1103/PhysRevLett.127.238002} {\bibfield  {journal} {\bibinfo  {journal} {Phys. Rev. Lett.}\ }\textbf {\bibinfo {volume} {127}},\ \bibinfo {pages} {238002} (\bibinfo {year} {2021})}\BibitemShut {NoStop}%
\bibitem [{\citenamefont {Tok{\'a}rov{\'a}}\ \emph {et~al.}(2021)\citenamefont {Tok{\'a}rov{\'a}}, \citenamefont {Sudalaiyadum~Perumal}, \citenamefont {Nayak}, \citenamefont {Shum}, \citenamefont {Ka{\v{s}}par}, \citenamefont {Rajendran}, \citenamefont {Mohammadi}, \citenamefont {Tremblay}, \citenamefont {Gaffney}, \citenamefont {Martel} \emph {et~al.}}]{Tokarova2021_PNAS}%
  \BibitemOpen
  \bibfield  {author} {\bibinfo {author} {\bibfnamefont {V.}~\bibnamefont {Tok{\'a}rov{\'a}}}, \bibinfo {author} {\bibfnamefont {A.}~\bibnamefont {Sudalaiyadum~Perumal}}, \bibinfo {author} {\bibfnamefont {M.}~\bibnamefont {Nayak}}, \bibinfo {author} {\bibfnamefont {H.}~\bibnamefont {Shum}}, \bibinfo {author} {\bibfnamefont {O.}~\bibnamefont {Ka{\v{s}}par}}, \bibinfo {author} {\bibfnamefont {K.}~\bibnamefont {Rajendran}}, \bibinfo {author} {\bibfnamefont {M.}~\bibnamefont {Mohammadi}}, \bibinfo {author} {\bibfnamefont {C.}~\bibnamefont {Tremblay}}, \bibinfo {author} {\bibfnamefont {E.~A.}\ \bibnamefont {Gaffney}}, \bibinfo {author} {\bibfnamefont {S.}~\bibnamefont {Martel}}, \emph {et~al.},\ }\href@noop {} {\bibfield  {journal} {\bibinfo  {journal} {Proceedings of the National Academy of Sciences}\ }\textbf {\bibinfo {volume} {118}},\ \bibinfo {pages} {e2013925118} (\bibinfo {year} {2021})}\BibitemShut {NoStop}%
\bibitem [{\citenamefont {Colburn}(1986)}]{Colburn1986}%
  \BibitemOpen
  \bibfield  {author} {\bibinfo {author} {\bibfnamefont {G.}~\bibnamefont {Colburn}},\ }\href@noop {} {\emph {\bibinfo {title} {The anatomy of the fallopian tube}}}\ (\bibinfo  {publisher} {Futura Publishing},\ \bibinfo {year} {1986})\BibitemShut {NoStop}%
\bibitem [{\citenamefont {Tung}\ \emph {et~al.}(2015{\natexlab{a}})\citenamefont {Tung}, \citenamefont {Hu}, \citenamefont {Fiore}, \citenamefont {Ardon}, \citenamefont {Hickman}, \citenamefont {Gilbert}, \citenamefont {Suarez},\ and\ \citenamefont {Wu}}]{tung_PNAS_2015}%
  \BibitemOpen
  \bibfield  {author} {\bibinfo {author} {\bibfnamefont {C.-k.}\ \bibnamefont {Tung}}, \bibinfo {author} {\bibfnamefont {L.}~\bibnamefont {Hu}}, \bibinfo {author} {\bibfnamefont {A.~G.}\ \bibnamefont {Fiore}}, \bibinfo {author} {\bibfnamefont {F.}~\bibnamefont {Ardon}}, \bibinfo {author} {\bibfnamefont {D.~G.}\ \bibnamefont {Hickman}}, \bibinfo {author} {\bibfnamefont {R.~O.}\ \bibnamefont {Gilbert}}, \bibinfo {author} {\bibfnamefont {S.~S.}\ \bibnamefont {Suarez}},\ and\ \bibinfo {author} {\bibfnamefont {M.}~\bibnamefont {Wu}},\ }\href@noop {} {\bibfield  {journal} {\bibinfo  {journal} {Proceedings of the National Academy of Sciences}\ }\textbf {\bibinfo {volume} {112}},\ \bibinfo {pages} {5431} (\bibinfo {year} {2015}{\natexlab{a}})}\BibitemShut {NoStop}%
\bibitem [{\citenamefont {Tung}\ \emph {et~al.}(2015{\natexlab{b}})\citenamefont {Tung}, \citenamefont {Ardon}, \citenamefont {Roy}, \citenamefont {Koch}, \citenamefont {Suarez},\ and\ \citenamefont {Wu}}]{Tung_PRL2015}%
  \BibitemOpen
  \bibfield  {author} {\bibinfo {author} {\bibfnamefont {C.-k.}\ \bibnamefont {Tung}}, \bibinfo {author} {\bibfnamefont {F.}~\bibnamefont {Ardon}}, \bibinfo {author} {\bibfnamefont {A.}~\bibnamefont {Roy}}, \bibinfo {author} {\bibfnamefont {D.~L.}\ \bibnamefont {Koch}}, \bibinfo {author} {\bibfnamefont {S.~S.}\ \bibnamefont {Suarez}},\ and\ \bibinfo {author} {\bibfnamefont {M.}~\bibnamefont {Wu}},\ }\href {https://doi.org/10.1103/PhysRevLett.114.108102} {\bibfield  {journal} {\bibinfo  {journal} {Phys. Rev. Lett.}\ }\textbf {\bibinfo {volume} {114}},\ \bibinfo {pages} {108102} (\bibinfo {year} {2015}{\natexlab{b}})}\BibitemShut {NoStop}%
\bibitem [{\citenamefont {Suarez}\ and\ \citenamefont {Wu}(2017)}]{suarez_sperm_2016}%
  \BibitemOpen
  \bibfield  {author} {\bibinfo {author} {\bibfnamefont {S.}~\bibnamefont {Suarez}}\ and\ \bibinfo {author} {\bibfnamefont {M.}~\bibnamefont {Wu}},\ }\href@noop {} {\bibfield  {journal} {\bibinfo  {journal} {MHR: Basic science of reproductive medicine}\ }\textbf {\bibinfo {volume} {23}},\ \bibinfo {pages} {227} (\bibinfo {year} {2017})}\BibitemShut {NoStop}%
\bibitem [{\citenamefont {Katz}(1974)}]{Katz1974}%
  \BibitemOpen
  \bibfield  {author} {\bibinfo {author} {\bibfnamefont {D.~F.}\ \bibnamefont {Katz}},\ }\href@noop {} {\bibfield  {journal} {\bibinfo  {journal} {J. Fluid Mech.}\ }\textbf {\bibinfo {volume} {64}},\ \bibinfo {pages} {33} (\bibinfo {year} {1974})}\BibitemShut {NoStop}%
\bibitem [{\citenamefont {M{\"{u}}nch}\ \emph {et~al.}(2016)\citenamefont {M{\"{u}}nch}, \citenamefont {Alizadehrad}, \citenamefont {Babu},\ and\ \citenamefont {Stark}}]{Munch2016}%
  \BibitemOpen
  \bibfield  {author} {\bibinfo {author} {\bibfnamefont {J.}~\bibnamefont {M{\"{u}}nch}}, \bibinfo {author} {\bibfnamefont {D.}~\bibnamefont {Alizadehrad}}, \bibinfo {author} {\bibfnamefont {S.}~\bibnamefont {Babu}},\ and\ \bibinfo {author} {\bibfnamefont {H.}~\bibnamefont {Stark}},\ }\href@noop {} {\bibfield  {journal} {\bibinfo  {journal} {Soft Matter}\ }\textbf {\bibinfo {volume} {12}},\ \bibinfo {pages} {7350} (\bibinfo {year} {2016})}\BibitemShut {NoStop}%
\bibitem [{\citenamefont {Taylor}(1952)}]{Taylor1952}%
  \BibitemOpen
  \bibfield  {author} {\bibinfo {author} {\bibfnamefont {G.~I.}\ \bibnamefont {Taylor}},\ }\href@noop {} {\bibfield  {journal} {\bibinfo  {journal} {Proc. Royal Soc. A}\ }\textbf {\bibinfo {volume} {211}},\ \bibinfo {pages} {225} (\bibinfo {year} {1952})}\BibitemShut {NoStop}%
\bibitem [{\citenamefont {Felderhof}(2010)}]{Felderhof2010}%
  \BibitemOpen
  \bibfield  {author} {\bibinfo {author} {\bibfnamefont {B.}~\bibnamefont {Felderhof}},\ }\href@noop {} {\bibfield  {journal} {\bibinfo  {journal} {Phys. Fluids}\ }\textbf {\bibinfo {volume} {22}},\ \bibinfo {pages} {113604} (\bibinfo {year} {2010})}\BibitemShut {NoStop}%
\bibitem [{\citenamefont {Evans}\ and\ \citenamefont {Lauga}(2010)}]{Evans2010}%
  \BibitemOpen
  \bibfield  {author} {\bibinfo {author} {\bibfnamefont {A.~A.}\ \bibnamefont {Evans}}\ and\ \bibinfo {author} {\bibfnamefont {E.}~\bibnamefont {Lauga}},\ }\href@noop {} {\bibfield  {journal} {\bibinfo  {journal} {Phys. Rev. E}\ }\textbf {\bibinfo {volume} {82}},\ \bibinfo {pages} {041915} (\bibinfo {year} {2010})}\BibitemShut {NoStop}%
\bibitem [{\citenamefont {Li}\ and\ \citenamefont {Ardekani}(2014)}]{Li2014}%
  \BibitemOpen
  \bibfield  {author} {\bibinfo {author} {\bibfnamefont {G.}~\bibnamefont {Li}}\ and\ \bibinfo {author} {\bibfnamefont {A.}~\bibnamefont {Ardekani}},\ }\href@noop {} {\bibfield  {journal} {\bibinfo  {journal} {Phys. Rev. E}\ }\textbf {\bibinfo {volume} {90}},\ \bibinfo {pages} {013010} (\bibinfo {year} {2014})}\BibitemShut {NoStop}%
\bibitem [{\citenamefont {Yazdi}\ \emph {et~al.}(2014)\citenamefont {Yazdi}, \citenamefont {Ardekani},\ and\ \citenamefont {Borhan}}]{Yazdi2014}%
  \BibitemOpen
  \bibfield  {author} {\bibinfo {author} {\bibfnamefont {S.}~\bibnamefont {Yazdi}}, \bibinfo {author} {\bibfnamefont {A.}~\bibnamefont {Ardekani}},\ and\ \bibinfo {author} {\bibfnamefont {A.}~\bibnamefont {Borhan}},\ }\href@noop {} {\bibfield  {journal} {\bibinfo  {journal} {Phys. Rev. E}\ }\textbf {\bibinfo {volume} {90}},\ \bibinfo {pages} {043002} (\bibinfo {year} {2014})}\BibitemShut {NoStop}%
\bibitem [{\citenamefont {Yazdi}\ \emph {et~al.}(2015)\citenamefont {Yazdi}, \citenamefont {Ardekani},\ and\ \citenamefont {Borhan}}]{Yazdi2015}%
  \BibitemOpen
  \bibfield  {author} {\bibinfo {author} {\bibfnamefont {S.}~\bibnamefont {Yazdi}}, \bibinfo {author} {\bibfnamefont {A.}~\bibnamefont {Ardekani}},\ and\ \bibinfo {author} {\bibfnamefont {A.}~\bibnamefont {Borhan}},\ }\href@noop {} {\bibfield  {journal} {\bibinfo  {journal} {J. Nonlinear Sci.}\ }\textbf {\bibinfo {volume} {25}},\ \bibinfo {pages} {1153} (\bibinfo {year} {2015})}\BibitemShut {NoStop}%
\bibitem [{\citenamefont {Schulman}\ \emph {et~al.}(2014)\citenamefont {Schulman}, \citenamefont {Backholm}, \citenamefont {Ryu},\ and\ \citenamefont {Dalnoki-Veress}}]{Schulman2014}%
  \BibitemOpen
  \bibfield  {author} {\bibinfo {author} {\bibfnamefont {R.}~\bibnamefont {Schulman}}, \bibinfo {author} {\bibfnamefont {M.}~\bibnamefont {Backholm}}, \bibinfo {author} {\bibfnamefont {W.}~\bibnamefont {Ryu}},\ and\ \bibinfo {author} {\bibfnamefont {K.}~\bibnamefont {Dalnoki-Veress}},\ }\href@noop {} {\bibfield  {journal} {\bibinfo  {journal} {Phys. Fluids}\ }\textbf {\bibinfo {volume} {26}},\ \bibinfo {pages} {101902} (\bibinfo {year} {2014})}\BibitemShut {NoStop}%
\bibitem [{\citenamefont {Bansil}\ \emph {et~al.}(1995)\citenamefont {Bansil}, \citenamefont {Stanley},\ and\ \citenamefont {Lamont}}]{bansil_ARPhys}%
  \BibitemOpen
  \bibfield  {author} {\bibinfo {author} {\bibfnamefont {R.}~\bibnamefont {Bansil}}, \bibinfo {author} {\bibfnamefont {E.}~\bibnamefont {Stanley}},\ and\ \bibinfo {author} {\bibfnamefont {J.~T.}\ \bibnamefont {Lamont}},\ }\href@noop {} {\bibfield  {journal} {\bibinfo  {journal} {Ann. Rev. Physiology}\ }\textbf {\bibinfo {volume} {57}},\ \bibinfo {pages} {635} (\bibinfo {year} {1995})}\BibitemShut {NoStop}%
\bibitem [{\citenamefont {Gaffney}\ \emph {et~al.}(2011)\citenamefont {Gaffney}, \citenamefont {Gadelha}, \citenamefont {Smith}, \citenamefont {Blake},\ and\ \citenamefont {Kirkman-Brown}}]{gaffney11}%
  \BibitemOpen
  \bibfield  {author} {\bibinfo {author} {\bibfnamefont {E.}~\bibnamefont {Gaffney}}, \bibinfo {author} {\bibfnamefont {H.}~\bibnamefont {Gadelha}}, \bibinfo {author} {\bibfnamefont {D.}~\bibnamefont {Smith}}, \bibinfo {author} {\bibfnamefont {J.}~\bibnamefont {Blake}},\ and\ \bibinfo {author} {\bibfnamefont {J.}~\bibnamefont {Kirkman-Brown}},\ }\href@noop {} {\bibfield  {journal} {\bibinfo  {journal} {Annu. Rev. Fluid Mech.}\ }\textbf {\bibinfo {volume} {43}},\ \bibinfo {pages} {501} (\bibinfo {year} {2011})}\BibitemShut {NoStop}%
\bibitem [{\citenamefont {Li}\ \emph {et~al.}(2014)\citenamefont {Li}, \citenamefont {Karimi},\ and\ \citenamefont {Ardekani}}]{Ardekani_RheoActa_2014}%
  \BibitemOpen
  \bibfield  {author} {\bibinfo {author} {\bibfnamefont {G.~J.}\ \bibnamefont {Li}}, \bibinfo {author} {\bibfnamefont {A.}~\bibnamefont {Karimi}},\ and\ \bibinfo {author} {\bibfnamefont {A.~M.}\ \bibnamefont {Ardekani}},\ }\href {https://doi.org/10.1007/s00397-014-0796-9} {\bibfield  {journal} {\bibinfo  {journal} {Rheo. Acta}\ }\textbf {\bibinfo {volume} {53}},\ \bibinfo {pages} {911} (\bibinfo {year} {2014})}\BibitemShut {NoStop}%
\bibitem [{\citenamefont {Elfring}\ and\ \citenamefont {Lauga}(2015)}]{Elfring_book2015}%
  \BibitemOpen
  \bibfield  {author} {\bibinfo {author} {\bibfnamefont {G.}~\bibnamefont {Elfring}}\ and\ \bibinfo {author} {\bibfnamefont {E.}~\bibnamefont {Lauga}},\ }in\ \href@noop {} {\emph {\bibinfo {booktitle} {Complex Fluids in Biological Systems}}},\ \bibinfo {editor} {edited by\ \bibinfo {editor} {\bibfnamefont {S.}~\bibnamefont {Spagnolie}}}\ (\bibinfo  {publisher} {Springer},\ \bibinfo {address} {New York, NY},\ \bibinfo {year} {2015})\BibitemShut {NoStop}%
\bibitem [{\citenamefont {Ives}\ and\ \citenamefont {Morozov}(2017)}]{Morozov_PoF2017}%
  \BibitemOpen
  \bibfield  {author} {\bibinfo {author} {\bibfnamefont {T.~R.}\ \bibnamefont {Ives}}\ and\ \bibinfo {author} {\bibfnamefont {A.}~\bibnamefont {Morozov}},\ }\href {https://doi.org/10.1063/1.4996839} {\bibfield  {journal} {\bibinfo  {journal} {Phys. Fluids}\ }\textbf {\bibinfo {volume} {29}},\ \bibinfo {pages} {121612} (\bibinfo {year} {2017})}\BibitemShut {NoStop}%
\bibitem [{\citenamefont {Sznitman}\ \emph {et~al.}(2010{\natexlab{c}})\citenamefont {Sznitman}, \citenamefont {Gupta}, \citenamefont {Hager}, \citenamefont {Arratia},\ and\ \citenamefont {Sznitman}}]{RSznitman2010}%
  \BibitemOpen
  \bibfield  {author} {\bibinfo {author} {\bibfnamefont {R.}~\bibnamefont {Sznitman}}, \bibinfo {author} {\bibfnamefont {M.}~\bibnamefont {Gupta}}, \bibinfo {author} {\bibfnamefont {G.}~\bibnamefont {Hager}}, \bibinfo {author} {\bibfnamefont {P.}~\bibnamefont {Arratia}},\ and\ \bibinfo {author} {\bibfnamefont {J.}~\bibnamefont {Sznitman}},\ }\href@noop {} {\bibfield  {journal} {\bibinfo  {journal} {PLoS ONE}\ }\textbf {\bibinfo {volume} {5}},\ \bibinfo {pages} {e11631} (\bibinfo {year} {2010}{\natexlab{c}})}\BibitemShut {NoStop}%
\bibitem [{\citenamefont {Brenner}(1974)}]{Brenner1974}%
  \BibitemOpen
  \bibfield  {author} {\bibinfo {author} {\bibfnamefont {S.}~\bibnamefont {Brenner}},\ }\href@noop {} {\bibfield  {journal} {\bibinfo  {journal} {Genetics}\ }\textbf {\bibinfo {volume} {77}},\ \bibinfo {pages} {71} (\bibinfo {year} {1974})}\BibitemShut {NoStop}%
\bibitem [{\citenamefont {Gagnon}\ \emph {et~al.}(2014{\natexlab{b}})\citenamefont {Gagnon}, \citenamefont {Keim},\ and\ \citenamefont {Arratia}}]{Gagnon2014}%
  \BibitemOpen
  \bibfield  {author} {\bibinfo {author} {\bibfnamefont {D.}~\bibnamefont {Gagnon}}, \bibinfo {author} {\bibfnamefont {N.}~\bibnamefont {Keim}},\ and\ \bibinfo {author} {\bibfnamefont {P.}~\bibnamefont {Arratia}},\ }\href@noop {} {\bibfield  {journal} {\bibinfo  {journal} {J. Fluid Mech.}\ }\textbf {\bibinfo {volume} {758}},\ \bibinfo {pages} {R3} (\bibinfo {year} {2014}{\natexlab{b}})}\BibitemShut {NoStop}%
\bibitem [{\citenamefont {Gagnon}\ and\ \citenamefont {Arratia}(2016)}]{Gagnon2016}%
  \BibitemOpen
  \bibfield  {author} {\bibinfo {author} {\bibfnamefont {D.}~\bibnamefont {Gagnon}}\ and\ \bibinfo {author} {\bibfnamefont {P.}~\bibnamefont {Arratia}},\ }\href@noop {} {\bibfield  {journal} {\bibinfo  {journal} {J. Fluid Mech.}\ }\textbf {\bibinfo {volume} {800}},\ \bibinfo {pages} {753} (\bibinfo {year} {2016})}\BibitemShut {NoStop}%
\bibitem [{\citenamefont {Gagnon}\ and\ \citenamefont {Montenegro-Johnson}(2018)}]{Gagnon2018}%
  \BibitemOpen
  \bibfield  {author} {\bibinfo {author} {\bibfnamefont {D.}~\bibnamefont {Gagnon}}\ and\ \bibinfo {author} {\bibfnamefont {T.}~\bibnamefont {Montenegro-Johnson}},\ }\href@noop {} {\bibfield  {journal} {\bibinfo  {journal} {ANZIAM J}\ }\textbf {\bibinfo {volume} {59}},\ \bibinfo {pages} {443} (\bibinfo {year} {2018})}\BibitemShut {NoStop}%
\bibitem [{\citenamefont {Majmudar}\ \emph {et~al.}(2012)\citenamefont {Majmudar}, \citenamefont {Keaveny}, \citenamefont {Zhang},\ and\ \citenamefont {Shelley}}]{majmudar2012experiments}%
  \BibitemOpen
  \bibfield  {author} {\bibinfo {author} {\bibfnamefont {T.}~\bibnamefont {Majmudar}}, \bibinfo {author} {\bibfnamefont {E.~E.}\ \bibnamefont {Keaveny}}, \bibinfo {author} {\bibfnamefont {J.}~\bibnamefont {Zhang}},\ and\ \bibinfo {author} {\bibfnamefont {M.~J.}\ \bibnamefont {Shelley}},\ }\href@noop {} {\bibfield  {journal} {\bibinfo  {journal} {J. Royal Soc. Interf.}\ }\textbf {\bibinfo {volume} {9}},\ \bibinfo {pages} {1809} (\bibinfo {year} {2012})}\BibitemShut {NoStop}%
\bibitem [{\citenamefont {Gray}\ and\ \citenamefont {Hancock}(1955)}]{Gray1955}%
  \BibitemOpen
  \bibfield  {author} {\bibinfo {author} {\bibfnamefont {J.}~\bibnamefont {Gray}}\ and\ \bibinfo {author} {\bibfnamefont {G.}~\bibnamefont {Hancock}},\ }\href@noop {} {\bibfield  {journal} {\bibinfo  {journal} {J. Exp. Biol.}\ }\textbf {\bibinfo {volume} {32}},\ \bibinfo {pages} {802} (\bibinfo {year} {1955})}\BibitemShut {NoStop}%
\bibitem [{\citenamefont {Thomases}\ and\ \citenamefont {Guy}(2017)}]{thomases2017role}%
  \BibitemOpen
  \bibfield  {author} {\bibinfo {author} {\bibfnamefont {B.}~\bibnamefont {Thomases}}\ and\ \bibinfo {author} {\bibfnamefont {R.~D.}\ \bibnamefont {Guy}},\ }\href@noop {} {\bibfield  {journal} {\bibinfo  {journal} {J. Fluid Mech.}\ }\textbf {\bibinfo {volume} {825}},\ \bibinfo {pages} {109} (\bibinfo {year} {2017})}\BibitemShut {NoStop}%
\bibitem [{\citenamefont {Bird}(1987)}]{Bird1987}%
  \BibitemOpen
  \bibfield  {author} {\bibinfo {author} {\bibfnamefont {R.}~\bibnamefont {Bird}},\ }\href@noop {} {\emph {\bibinfo {title} {Dynamics of polymeric liquids}}}\ (\bibinfo  {publisher} {Wiley},\ \bibinfo {address} {New York},\ \bibinfo {year} {1987})\BibitemShut {NoStop}%
\bibitem [{\citenamefont {Montenegro-Johnson}\ \emph {et~al.}(2016)\citenamefont {Montenegro-Johnson}, \citenamefont {Gagnon}, \citenamefont {Arratia},\ and\ \citenamefont {Lauga}}]{Montenegro2016}%
  \BibitemOpen
  \bibfield  {author} {\bibinfo {author} {\bibfnamefont {T.}~\bibnamefont {Montenegro-Johnson}}, \bibinfo {author} {\bibfnamefont {D.}~\bibnamefont {Gagnon}}, \bibinfo {author} {\bibfnamefont {P.}~\bibnamefont {Arratia}},\ and\ \bibinfo {author} {\bibfnamefont {E.}~\bibnamefont {Lauga}},\ }\href@noop {} {\bibfield  {journal} {\bibinfo  {journal} {Phys. Rev. Fluids}\ }\textbf {\bibinfo {volume} {1}},\ \bibinfo {pages} {053202} (\bibinfo {year} {2016})}\BibitemShut {NoStop}%
\end{thebibliography}%

\end{document}